\documentclass[preprintnumbers,prd,showpacs,floatfix,preprintnumbers,
superscriptaddress,nofootinbib,twocolumn]{revtex4-1}
\pdfoutput=1
\usepackage{graphicx,epsfig}
\usepackage{amsmath}
\usepackage {amssymb}
\usepackage {longtable}
\usepackage{multirow}

\usepackage{dcolumn}
\usepackage{bm}
\usepackage{amsfonts}
\usepackage{subfigure}
\usepackage{color}

\usepackage{relsize}

\def\be{\begin{equation}}
\def\ee{\end{equation}}
\def\beq{\begin{eqnarray}}
\def\eeq{\end{eqnarray}}

\newcommand{\al}{\alpha}
\newcommand{\bt}{\beta}
\newcommand{\R}{\mathcal{R}}

\begin{document}

\title{Cosmological solutions in generalized hybrid
metric-Palatini gravity}

\author{Jo\~{a}o L. Rosa}
\email{joaoluis92@gmail.com}
\affiliation{Centro Multidisciplinar de Astrof\'{\i}sica - CENTRA,
Departamento de F\'{\i}sica,
Instituto Superior T\'{e}cnico - IST,
Universidade de Lisboa - UL,
Avenida Rovisco Pais 1, 1049-001, Lisbon, Portugal}

\author{Sante Carloni}
\email{sante.carloni@gmail.com}
\affiliation{Centro Multidisciplinar de Astrof\'{\i}sica - CENTRA,
Departamento de F\'{\i}sica,
Instituto Superior T\'{e}cnico - IST,
Universidade de Lisboa - UL,
Avenida Rovisco Pais 1, 1049-001, Lisbon, Portugal}

\author{Jos\'{e} P. S. Lemos}
\email{joselemos@ist.utl.pt}
\affiliation{Centro Multidisciplinar de Astrof\'{\i}sica - CENTRA,
Departamento de F\'{\i}sica,
Instituto Superior T\'{e}cnico - IST,
Universidade de Lisboa - UL,
Avenida Rovisco Pais 1, 1049-001, Lisbon, Portugal}

\author{Francisco S. N. Lobo}
\email{fslobo@fc.ul.pt}
\affiliation{Instituto de Astrof\'{\i}sica e Ci\^{e}ncias do Espa\c{c}o, Faculdade de
Ci\^encias, Universidade de Lisboa, Edif\'{\i}cio C8, Campo Grande,
P-1749-016 Lisbon, Portugal}

\date{\today}

\begin{abstract} 
We construct exact solutions representing a Friedmann-Lema\^itre-Robsertson-Walker (FLRW) universe in a generalized hybrid metric-Palatini theory. By writing the gravitational action in a scalar-tensor representation, the new solutions are obtained by either making an ansatz on the scale factor or on the effective potential. Among other relevant results, we show that it is possible to obtain exponentially expanding solutions for flat universes even when the cosmology is not purely vacuum.  We then derive the classes of actions for the original theory which generate these solutions.

\end{abstract}


\maketitle

%%%%%%%%%%%%%%%%%%%%%%%%%%%%%%%%%%%%%%%%%%%%%%%%%%%%%%%%%%%%%%%%%%%%
\section{Introduction}\label{Introduction}
%%%%%%%%%%%%%%%%%%%%%%%%%%%%%%%%%%%%%%%%%%%%%%%%%%%%%%%%%%%%%%%%%%%%

The late-time cosmic accelerated expansion
\cite{Perlmutter:1998np,Riess:1998cb} has posed important and
challenging problems to theoretical cosmology. Although the standard
model of cosmology has favored dark energy models
\cite{Copeland:2006wr} as fundamental candidates responsible for the
accelerated cosmic expansion, it is also viable that this expansion is
due to modifications of general relativity
\cite{Clifton:2011jh,Nojiri:2010wj}, which introduce new degrees of
freedom to the gravitational sector itself.  Indeed, the phenomenology
of $f(R)$ gravity, where $R$ is the metric Ricci curvature scale and $f$ a
general function,
has been scrutinized motivated by the possibility to
account for the self-accelerated cosmic expansion without invoking
dark energy sources \cite{Sotiriou:2008rp,DeFelice:2010aj}. Besides,
this kind of modified gravity is capable of addressing the dynamics of
several self-gravitating systems alternatively to the presence of dark
matter \cite{Boehmer:2007kx,Bohmer:2007fh}.
In this approach of $f(R)$ gravity, and using its equivalent scalar-tensor
representation,
one can show that in order to satisfy local, i.e., solar system, observational constraints
a large mass of the scalar field is required, 
which scales with the curvature through the chameleon mechanism \cite{Khoury:2003aq,Khoury:2003rn}.
In turn, this has undesirable effects at cosmological scales.
There are other drawbacks in these models,
see \cite{Sotiriou:2008rp,DeFelice:2010aj,Boehmer:2007kx,Bohmer:2007fh}.
A Palatini version of the  $f(R)$ gravity
theory, where the connection rather than the metric represents the fundamental
gravitational field, has been proposed \cite{Olmo:2011uz}, and it has been
established that it has
interesting features but the manifest deficiencies and
downsides of the metric $f(R)$ gravity also appear 
\cite{Olmo:2011uz}.

A hybrid combination of the two versions of the  $f(R)$ gravity theories,
containing elements from both of
the two formalisms, i.e., a hybrid metric-Palatini theory,
consists of adding an $f({\cal R})$ term
constructed \`{a} la Palatini to the
Einstein-Hilbert Lagrangian \cite{Harko:2011nh}.
It turns out to be very successful in accounting for the observed
phenomenology and is able to avoid some of the shortcommings of the
original approaches \cite{Capozziello:2013uya}.
In the scalar-tensor representation of the hybrid metric-Palatini
theory there is a long-range light mass scalar field, which is able to
modify, in a way consistent with the observations, the cosmological
and galactic dynamics, but leaves the solar system unaffected
\cite{Capozziello:2012ny,Capozziello:2012qt}. This light scalar field
thus allows to evade the screening chameleon mechanism.
In addition, absence of instabilities in perturbations were
also verified in this model \cite{Koivisto:2013kwa}.

The theory was further developed into a generalized
hybrid metric-Palatini theory, in which a general function, $f\left(R,\mathcal R\right)$,
was postulated depending 
on
both the metric and Palatini curvature scalars, $R$ and $\mathcal R$,
respectively
\cite{Tamanini:2013ltp}.
Of course, the positive
aspects of the initial hybrid metric-Palatini theory
are preserved, namely, the
consistency with
cosmological and galactic dynamics, and correct solar system
tests
\cite{Carloni:2015bua,Capozziello:2013yha}, see \cite{Capozziello:2015lza} for a review.
There are other developments.
Considering linear homogeneous perturbations, the stability regions of the Einstein static universe were analyzed, and it was shown that a large class of stable solutions exists \cite{Boehmer:2013oxa}.
In addition,
the full set of linearized evolution equations for the perturbed potentials, in the Newtonian and synchronous gauges, were derived. It was concluded that the main deviations from general relativity arise in the distant past, with an oscillatory signature in the ratio between the Newtonian potentials \cite{Lima:2014aza}. Furthermore, using specific models and a combination of cosmic microwave background, supernovae and baryonic acoustic oscillations background data, it was shown that the model's free parameters are in agreement with the observational constraints \cite{Lima:2015nma,Lima:2015nma,Lima:2016npg}.
It was also shown in this theory that the
initial value problem can always be well-formulated and be well-posed
depending on the adopted matter sources \cite{Capozziello:2013gza}.

The understanding and the building of solutions in complex theories like
the generalized hybrid
metric-Palatini gravity, and its scalar-tensor representation,  is a difficult task. 
Among other methods to find solutions, 
the reconstruction technique method has been often valuable
in this search.
This method was first employed in
\cite{LM} in order to select the form for the inflation potential able
to resolve the open problems of the inflationary paradigm, e.g., the
graceful exit. Also other elegant attempts were made to
generalize this approach to nonminimally coupled scalar-tensor
theories with a single scalar field \cite{Ellis:1990wsa}.

In this work, we aim to find cosmological solutions in the generalized
hybrid metric-Palatini gravity proposed in \cite{Tamanini:2013ltp}
through the use of its scalar-tensor representation.  We will then
devise a reconstruction technique algorithm for scalar-tensor theories
in a Friedmann-Lema\^itre-Robsertson-Walker (FLRW) universe and we
will show that the generalized theory through its scalar-tensor
representation provides a very rich structure in astrophysical and
cosmological applications.  In particular, we will find that the
cosmology of the generalized hybrid metric-Palatini gravity can differ
in subtle ways from both general relativity and $f(R)$ gravity. In
principle these differences can be used in combination with
observational data to verify the viability of this class of theories.

This paper is organized in the following manner: In Sec.~\ref{sec:action}, we present the formalism of the generalized hybrid metric-Palatini gravity, and express the action in the scalar-tensor representation. In Sec.~\ref{secIII}, we consider the cosmological dynamics of FLRW spacetimes, using the scalar-tensor representation of the hybrid theory. In Sec.~\ref{secIV}, we obtain, from the
solutions found in the scalar-tensor representation, the specific forms for $f\left(R,\cal{R}\right)$. In Sec.~\ref{conclusion},  we set out our conclusions.

%%%%%%%%%%%%%%%%%%%%%%%%%%%%%%%%%%%%%%%%%%%%%%%%%%%%%%%%%%%%%%%%%%%%
\section{Generalized hybrid gravity: Formalism}\label{sec:action}
%%%%%%%%%%%%%%%%%%%%%%%%%%%%%%%%%%%%%%%%%%%%%%%%%%%%%%%%%%%%%%%%%%%%

%%%%%%%%%%%%%%%%%%%%%%%%%%%%%%%%%%%%%%%%%%%%%%%%%%%%%%%%%%%%%%%%%%%%
\subsection{Action and field equations for the generalized hybrid gravity}
%%%%%%%%%%%%%%%%%%%%%%%%%%%%%%%%%%%%%%%%%%%%%%%%%%%%%%%%%%%%%%%%%%%%

Consider the action of the generalized hybrid metric-Palatini modified theory of gravity, given by (the velocity of light is set to one)
\be\label{genac}
S=\frac{1}{2\kappa^2}\int_\Omega\sqrt{-g}f\left(R,\cal{R}\right)d^4x+S_m(g_{ab}, \chi),
\ee
where $\kappa\equiv 8\pi G$, 
$G$ is the gravitational constant, $g$ is the determinant of the metric
$g_{ab}$, $f$ is a  function of $R$ and $\mathcal{R}$, and $S_m$ is the matter action,  in which matter is minimally coupled to the metric $g_{ab}$, and $\chi$ collectively denotes the matter fields. $R$ is the metric Ricci scalar and 
$\mathcal{R}\equiv g_{ab}\mathcal{R}^{ab}$ is the Palatini scalar
curvature, with $\mathcal{R}^{ab}$
being
defined in terms of an independent connection $\hat\Gamma^c_{ab}$ as, 
\be
\mathcal{R}_{ab}=\partial_c\hat\Gamma^c_{ab}-\partial_b
\hat\Gamma^c_{ac}+\hat\Gamma^c_{cd}\hat\Gamma^d_{ab}-\hat\Gamma^c_{ad}
\hat\Gamma^d_{cb}\,.
\label{RicciPalatini}
\ee
One can also use the additional independent connection as a building block to construct higher order curvature invariants, in the action 
$
S = \frac{1}{2\kappa^2} \int {\rm d}^4x \sqrt{-g}\, f( R,\R,{\cal{Q}}_H )
$,
where the term ${\cal{Q}}_H$, can take the following forms
${\cal{R}}^{\mu\nu} {\cal{R}}_{\mu\nu}$, $R^{\mu\nu}{\cal{R}}_{\mu\nu}$, ${\cal{R}}^{\mu\nu\al\bt}{\cal{R}}_{\mu\nu\al\bt}$, $R^{\mu\nu\al\bt}{\cal{R}}_{\mu\nu \al\bt}$, ${\cal{R}} R$, etc. However we will not consider these cases here. Relative to the presence of Ostrogradski instabilities, in the hybrid theory, like in $f(R)$-gravity,  it is possible to avoid the problem by allowing a separation of the additional degrees of freedom into a harmless scalar degree of freedom \cite{Sotiriou:2008rp,DeFelice:2010aj}. In fact, it turns out that this feature is a similar exception in the larger space of metric-affine theories, where a generic theory is plagued by ghosts, superluminalities and other unphysical degrees of freedom \cite{Koivisto:2013kwa}.

Varying the action (\ref{genac}) with respect to the metric $g_{ab}$ and the independent connection, yields the following field equations
\beq
\frac{\partial f}{\partial R}R_{ab}+\frac{\partial f}{\partial 
\mathcal{R}}\mathcal{R}_{ab}-\frac{1}{2}g_{ab}f\left(R,\cal{R}\right)
  \nonumber  \\
-\left(\nabla_a\nabla_b-g_{ab}\Box\right)\frac{\partial f}{\partial R}=\kappa^2 T_{ab},
\eeq
and
\be
\hat\nabla_c\left(\sqrt{-g}\frac{\partial f}{\partial \cal{R}}g^{ab}\right)=0,
  \label{indconnnection}
\ee
respectively. The equation of motion (\ref{indconnnection}) implies that the independent connection is the Levi-Civita connection of a 
new metric tensor $h_{ab}$ which is conformally related to $g_{ab}$ by the relation 
\be\label{hgf}
h_{ab}=g_{ab} \frac{\partial f}{\partial \cal{R}}\,.
\ee
The independent connection can then be written in terms of the new metric as
\be
\hat\Gamma^a_{bc}=\frac{1}{2}h^{ad}\left(\partial_b h_{dc}+
\partial_c h_{bd}-\partial_d h_{bc}\right).
\ee
Thus, the new metric $h_{ab}$ is an auxiliary metric related to the independent connection, that was used to define the Palatini tensor, given by Eq. (\ref{RicciPalatini}). We emphasize that matter is coupled to the physical metric $g_{ab}$, so that only the Levi-Civita connection $\Gamma^a_{bc}(g)$  should be used in the geodesic equation applied to the metric-Palatini theory. 
Note that since the matter action $S_m$ does not depend on the independent connection $\hat\Gamma$, the equation of motion for this connection is independent of the stress-energy tensor, whereas the same does not happen to the equation of motion of the metric $g_{ab}$.

%%%%%%%%%%%%%%%%%%%%%%%%%%%%%%%%%%%%%%%%%%%%%%%%%%%%%%%%%%%%%%%%%%%%
\subsection{Scalar-tensor representation}
%%%%%%%%%%%%%%%%%%%%%%%%%%%%%%%%%%%%%%%%%%%%%%%%%%%%%%%%%%%%%%%%%%%%

It is useful to express the action~\eqref{genac} in a scalar-tensor representation. This can be achieved by considering an action with two auxiliary fields, $\alpha$ and $\beta$, respectively, in the following form
\beq
S=\frac{1}{2\kappa^2}\int_\Omega \sqrt{-g}\;\Big[f\left(\alpha,\beta\right)+\frac{\partial f}{\partial \alpha}\left(R-\alpha\right)
    \nonumber \\
+\frac{\partial f}{\partial\beta}\left(\cal{R}-\beta\right)\Big]d^4x+S_m.
\label{gensca}
\eeq
Putting $\alpha=R$ and $\beta=\mathcal{R}$ we recover action~\eqref{genac}.
To proceed, we define two scalar fields as 
\be\label{phidef}
\varphi=\frac{\partial f}{\partial\alpha}\,,
\ee
\be\label{psidef}
\psi=-\frac{\partial f}{\partial\beta}\,,
\ee
where the negative sign in Eq.~\eqref{psidef}
is imposed to guarantee a positive kinetic energy for the scalar field.
Defining a potential $V$ as
\be\label{potdef}
V\left(\varphi,\psi\right)=-f\left(\alpha,\beta\right)+\varphi\alpha-\psi\beta\,,
\ee
the  action equivalent to (\ref{gensca}) is of the form
\be
S=\frac{1}{2\kappa^2}\int_\Omega \sqrt{-g}\left[\varphi R-\psi\mathcal{R}-V\left(\varphi,\psi\right)\right]d^4x+S_m\,,
  \label{action3}
\ee
where the variation of the matter action $S_m$ with respect to the metric $g_{ab}$
yields the stress-energy tensor $T_{ab}$. 

It is important to stress that in this representation the scalar fields are not actually matter fields but only alternative representations of the curvatures $R$ and $\mathcal R$, and consequently they are allowed to assume values, such as negative ones, which would not be physically valid.Also, since we imposed the signs on the definitions of $\psi$ and $\varphi$ to guarantee that the kinetic energies are positive, then we should not worry about ghost instabilities. This range of values corresponds in the generalized hybrid gravity picture to some specific behavior of the derivatives of the function $f$ in the action.  That said, it should also be clear that zeros for the scalar fields are problematic points in terms of the relation between the representations. In the following, the solutions we will obtain are only valid in the interval between the zeros of the scalar fields. In addition, it should be remarked that the scalar field representation will only be meaningful provided that $R$  and $\mathcal{R}$ can be expressed in terms of the scalar fields. This is not the case for all forms of the function $f$. In the next sections we will show some examples in which the consequences of these limitations become important.

Taking into account that $h_{ab}=-\psi\, g_{ab}$, see Eqs. (\ref{hgf}) and (\ref{psidef}), we have
\be\label{confrt}
\mathcal{R}=R+\frac{3}{\psi^2}\partial^a \psi\partial_a \psi-\frac{3}{\psi}\Box\psi \,.
\ee
Thus, we can replace $\cal{R}$ in the action (\ref{action3}), 
to obtain
\beq\label{genacts2}
S=\frac{1}{2\kappa^2}\int_\Omega \sqrt{-g}\Big[\left(\varphi-\psi\right) R-\frac{3}{2\psi}\partial^a\psi\partial_a\psi
     \nonumber  \\
-V\left(\varphi,\psi\right)\big]d^4x+S_m.
\eeq

Varying the previous action, Eq.~(\ref{genacts2}), with respect to the metric $g_{ab}$ and the scalar fields $\varphi$ and $\psi$, and rearranging terms, we obtain the following equations of motion
\beq
&&\left(\varphi-\psi\right) G_{ab}=\kappa^2T_{ab}+\left(\nabla_a\nabla_b-g_{ab}\Box\right)\left(\varphi-\psi\right)
   \nonumber  \\
&&\ +\frac{3}{2\psi}\partial_a\psi\partial_b\psi+\left(\frac{1}{2}V+\frac{3}{4\psi}\partial^c\psi\partial_c\psi\right)g_{ab}\,,
\label{genein2}
\eeq
\be\label{genphi}
\Box\varphi+\frac{1}{3}\left(2V-\psi V_\psi-\varphi V_\varphi\right)=\frac{\kappa^2T}{3} \,,
\ee
\be
\Box\psi-\frac{1}{2\psi}\partial^a\psi\partial_a\psi-\frac{\psi}{3}\left(V_\varphi+V_\psi\right)=0\,,
\label{genpsi}
\ee
respectively, where $V_\varphi$ and $V_\psi$ are defined 
by
\be
V_\varphi\equiv \frac{\partial V}{\partial\varphi}\,, \qquad
V_\psi\equiv \frac{\partial V}{\partial\psi}\,,
\ee
respectively.

The dynamical equation for $\varphi$, Eq.~(\ref{genphi}), can be obtained in the following manner. The variation of the action with respect to
 $\varphi$ gives 
 \be\label{rvphi}
 R=V_{\varphi}. 
 \ee
 Then, taking the trace of Eq.~(\ref{genein2})
 and inserting $R=V_{\varphi}$ gives Eq.~(\ref{genphi}).
 The latter equation
shows an important difference between the two scalar fields, namely, $\varphi$ is coupled to matter whereas $\psi$, given through Eq.~(\ref{genpsi}), is not. This fact will have important consequences when we will look for cosmological solutions in which  matter is present.

%%%%%%%%%%%%%%%%%%%%%%%%%%%%%%%%%%%%%%%%%%%%%%%%%%%%%%%%%%%%%%%%%%%%
\section{Cosmological equations and solutions for the generalized hybrid gravity in the scalar-tensor representation}\label{secIII}
%%%%%%%%%%%%%%%%%%%%%%%%%%%%%%%%%%%%%%%%%%%%%%%%%%%%%%%%%%%%%%%%%%%%

%%%%%%%%%%%%%%%%%%%%%%%%%%%%%%%%%%%%%%%%%%%%%%%%%%%%%%%%%%%%%%%%%%%%
\subsection{Cosmological equations in the scalar-tensor representation}\label{seceqsstr}
%%%%%%%%%%%%%%%%%%%%%%%%%%%%%%%%%%%%%%%%%%%%%%%%%%%%%%%%%%%%%%%%%%%%

In this section, we consider the FLRW spacetime with the spatial curvature parameter $k$, where $k$ can assume three values, $k=-1,0,1$. In spherical coordinates
$(t,r,\theta,\phi)$
the line element can be written as
\be\label{FLRW}
ds^2=-dt^2+a^2\left(t\right)\left[\frac{dr^2}{1-kr^2}+r^2d\theta^2+r^2\sin^2\theta d\phi^2\right],
\ee
where $a(t)$ is the scale factor.
We also assume that the cosmological scalar fields are time-dependent, $\varphi=\varphi\left(t\right)$, $\psi=\psi\left(t\right)$, and that the matter is described by a perfect fluid,
\be\label{perffluid}
T^a_b={\rm diag}\left(-\rho,p,p,p\right)\,,
\ee
so that the trace is given by
\be\label{tabtrace}
T=T^a_a=-\rho+3p\,,
\ee
with $\rho$ and $p$ being the energy density and the pressure of the fluid, respectively, and both are assumed to depend solely on $t$.

Then the modified cosmological equations for this case can be obtained 
by computing directly the two independent components of
the gravitational field equation, Eq.~\eqref{genein2}, i.e., the Friedmann equation
\beq\label{genH}
3\left(\frac{\dot a}{a}\right)^2+\frac{3k}{a^2}=\frac{1}{\varphi-\psi}\Bigg[\kappa^2\rho+\frac{V}{2}
   \nonumber  \\
+3\left(\frac{\dot\psi^2}{4\psi}-\left(\frac{\dot a}{a}\right)\left(\dot\varphi-\dot\psi\right)\right)\Bigg],  
\eeq
and the Raychaudhuri equation
\beq
\label{gendH}
2\frac{d}{dt} {\left(\frac{\dot a}{a}\right)}-\frac{2k}{a^2}=\frac{1}{\varphi-\psi}\Bigg[-\kappa^2\left(\rho+p\right)-\frac{3\dot\psi^2}{2\psi}
     \nonumber  \\
+\left(\frac{\dot a}{a}\right)\left(\dot\varphi-\dot\psi\right)-\left(\ddot\varphi-\ddot\psi\right)\Bigg]\,,
\eeq
respectively, where a dot means differentiation with respect
to $t$, $\;\dot{}\equiv d/dt$.

The evolution equations
for the scalar fields, Eqs.~\eqref{genphi} and 
\eqref{genpsi}, take the following forms 
\be
\label{kgphi0}
\ddot\varphi+3\left(\frac{\dot a}{a}\right)\dot\varphi-\frac{1}{3}\left[2V-\psi V_\psi-\varphi V_\varphi\right]=-\frac{\kappa^2T}{3}\,,
\ee
\be\label{kgpsi0}
\ddot\psi+3\left(\frac{\dot a}{a}\right)\dot\psi-\frac{\dot\psi^2}{2\psi}+\frac{\psi}{3}\left(V_\varphi+V_\psi\right)=0\,,
\ee
respectively. 

Now, an equation for the potential can be obtained by multiplying Eq.~\eqref{gendH} by $3/2$, summing it to Eq.~\eqref{genH}, and using Eqs.~\eqref{kgphi0} and~\eqref{kgpsi0} to cancel the terms $\ddot\varphi$ and $\ddot\psi$, similarly to the method performed in \cite{Ellis:1990wsa}. The result is as follows,
\be\label{modF1}
V_\varphi=6\left[
{\frac{d}{dt} \left(\frac{\dot a}{a}\right)
} +2
\left(\frac{\dot a}{a}\right)^2+\frac{k}{a^2}
\right]\,.
\ee

Finally, the equations to describe the evolution of matter are an equation of state of the form
\be\label{eos}
p=w\rho, 
\ee
where $w$ is a parameter without units, 
and the conservation law for the stress-energy tensor $\nabla_a T^{ab}=0$, which becomes, using the equation of state~\eqref{eos},
\be\label{econs}
\dot\rho=-3\,\frac{\dot a}{a}\,\rho\left(1+w\right).
\ee

We have now a system of five independent equations, namely~\eqref{kgphi0}--\eqref{econs}, that
we have to solve for the seven independent variables $k,a, \rho, p, V, \varphi$, and $\psi$.
We will choose some specific geometry, i.e., we choose $k$.
We can then impose one extra constraint to the system in order for it to have a unique solution. In the following, we shall impose different constraints and obtain their respective solutions.

It is also useful to define
the Hubble function or parameter  $H$ as usual
\be\label{hubblep}
H=\frac{\dot a}{a}\,,
\ee
which is a function of $t$. With Eq.~\eqref{hubblep}
we can trade $a(t)$ for $H(t)$ and vice versa.

%%%%%%%%%%%%%%%%%%%%%%%%%%%%%%%%%%%%%%%%%%%%%%%%%%%%%%%%%%%%%%%%%%%%
\subsection{Cosmological solutions in the scalar-tensor representation}\label{secIVb}
%%%%%%%%%%%%%%%%%%%%%%%%%%%%%%%%%%%%%%%%%%%%%%%%%%%%%%%%%%%%%%%%%%%%

%%%%%%%%%%%%%%%%%%%%%%%%%%%%%%%%%%%%%%%%%%%%%%%%%%%%%%%%%%%%%%%%%%%%
\subsubsection{The de Sitter solution}\label{method1}
%%%%%%%%%%%%%%%%%%%%%%%%%%%%%%%%%%%%%%%%%%%%%%%%%%%%%%%%%%%%%%%%%%%%

In this section, we try to find a solution of the form of a de Sitter expansion.
To do so
we
set the matter component to be vacuum, that is, $\rho=p=0$.
%\be
%\rho=0\,,
%\ee
%\be
%p=0\,.
%\ee
This choice means that we actually do not need Eqs.~\eqref{eos} and~\eqref{econs} to solve the system.

\begin{figure}[h!]
\centering \includegraphics[scale=0.65]{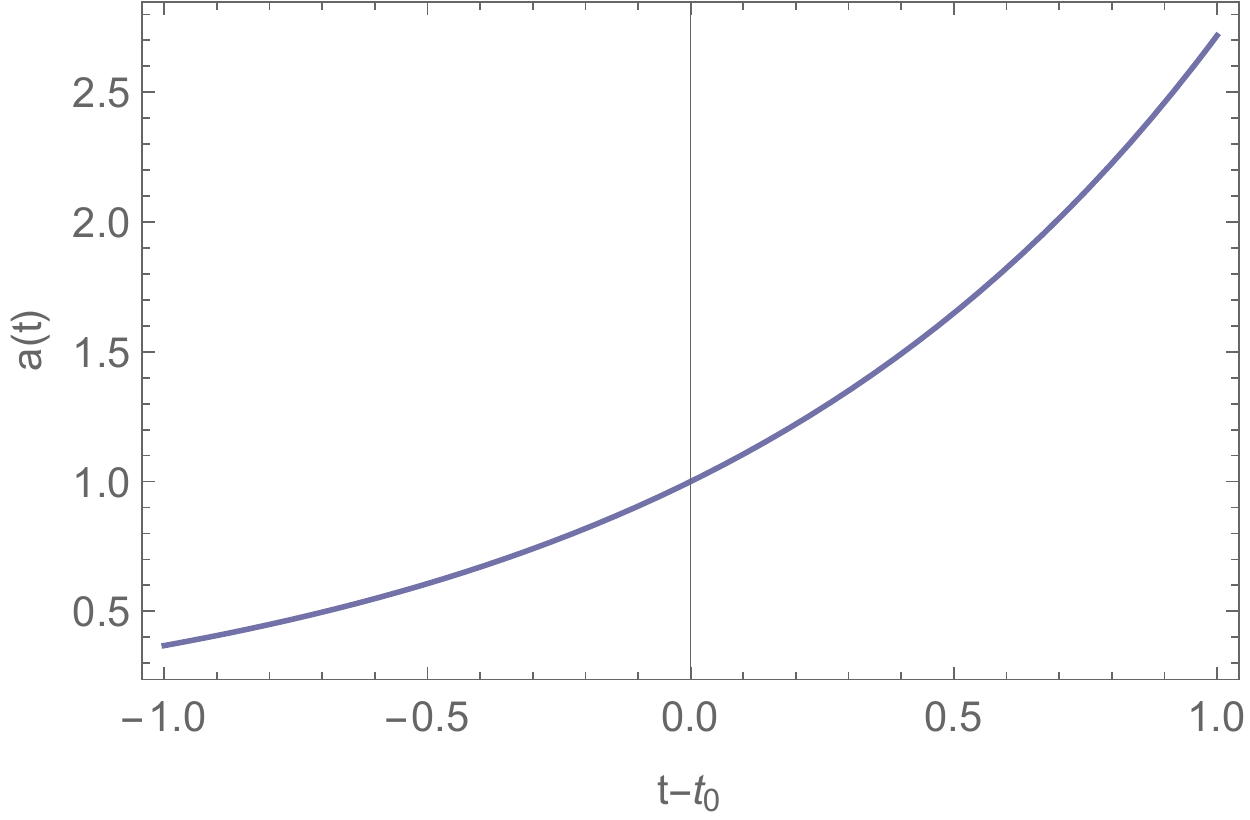}
\caption{Plot of the scale factor $a(t)$ for the de Sitter case, given in Eq.~\eqref{scalefactordes1},
for the values $a_0=1$,  $\Lambda=1$.}
\label{fig:a1}
\end{figure}

We also choose a flat universe, i.e., $k=0$
%\be
%k=0\,,
%\ee
and a scale factor described by 
\be\label{scalefactordes1}
a\left(t\right)=Ae^{\sqrt\Lambda\, \left(t-t_0\right)}\,,
\ee
where $A$ and $\Lambda$ are free constants, and
$\Lambda$ can be seen as some cosmological constant. This scale factor is plotted in Fig. \ref{fig:a1}.

\begin{figure}[h!]
\centering \includegraphics[scale=0.65]{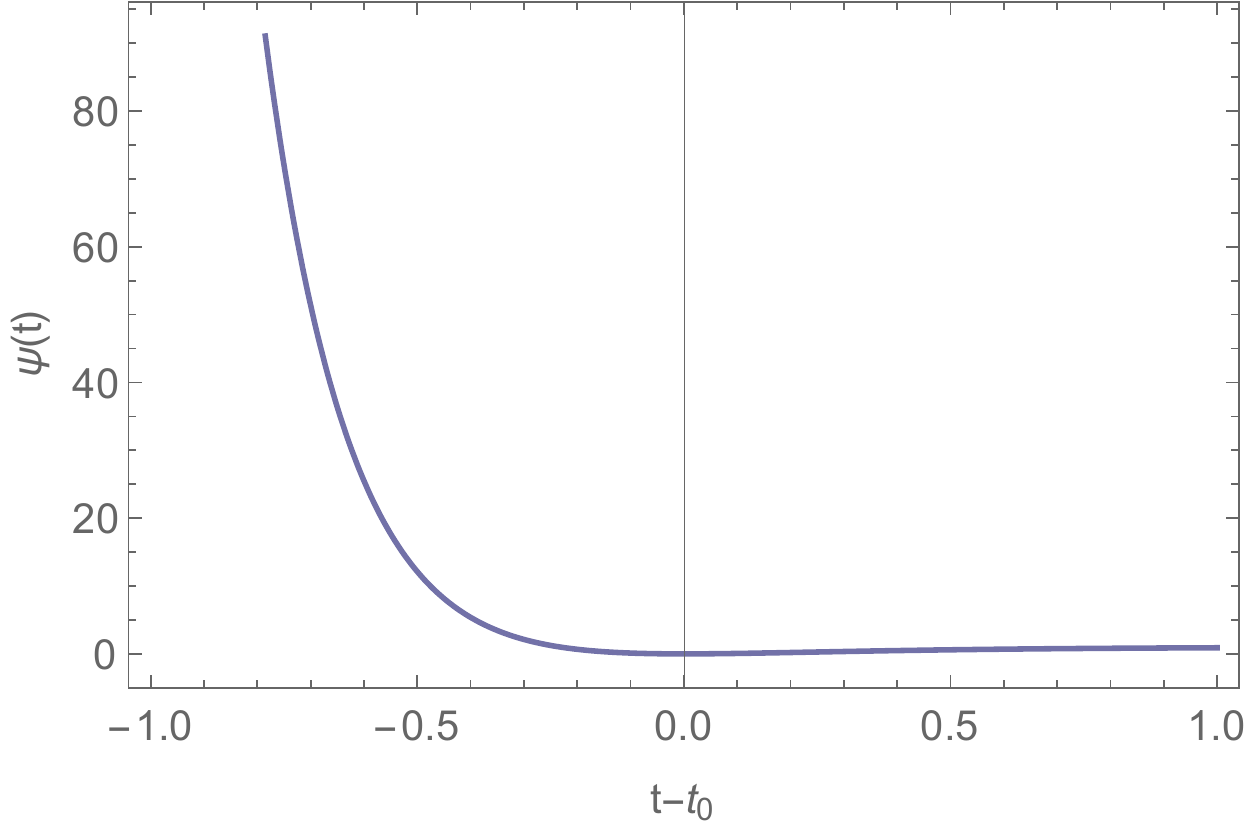}
\caption{Plot of the scalar field $\psi(t)$ for the de Sitter case, given in Eq.~\eqref{solpsi},
for the values $\psi_0=\Lambda=1$, and $t_0=0$.}
\label{fig:psi1}
\end{figure}

\begin{figure}[h!]
\centering \includegraphics[scale=0.65]{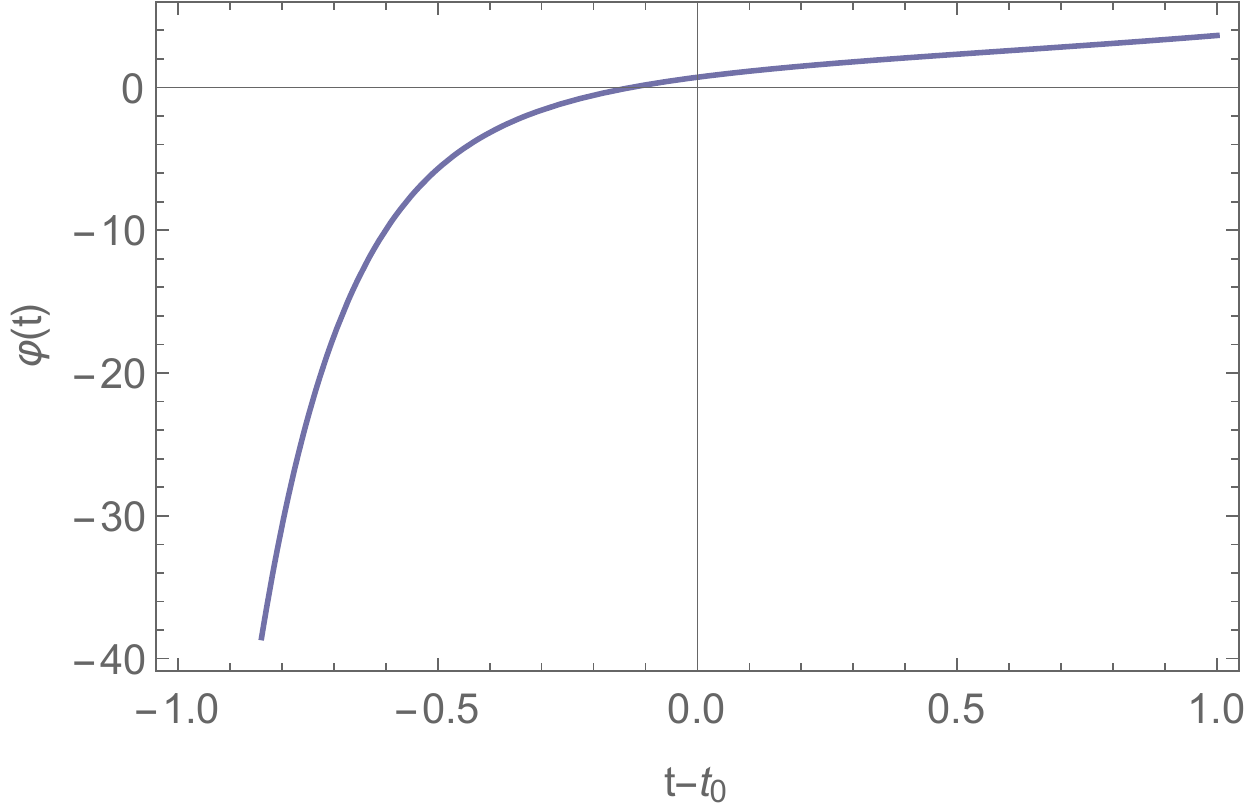}
\caption{Plot of the scalar field $\phi(t)$ for the de Sitter case, given in Eq.~\eqref{solvarphiades1},
for the values $\phi_0=\phi_1=\Lambda=1$, and $t_0=0$.}
\label{fig:phi1}
\end{figure}

With these assumptions, Eq.~\eqref{modF1} becomes
\be\label{solV1}
V_\varphi=12\Lambda.
\ee
This equation can be directly integrated with respect to $\varphi$ to obtain a potential of the form
\be
V\left(\varphi,\psi\right)=12\Lambda\varphi+b\left(\psi\right).
\ee
where $b\left(\psi\right)$ is an arbitrary function on $\psi$ which arises from the fact that the potential is a function of both fields. 
These results leave Eqs.~\eqref{kgphi0} and~\eqref{kgpsi0} as
\be
\ddot\varphi+3\sqrt\Lambda\,\dot\varphi-\frac{1}{3}\left[12\Lambda\,\varphi+2b\left(\psi\right)-\psi b'\left(\psi\right)\right]=0\,,
\label{dotphi}
\ee
\be
\ddot\psi+3\sqrt\Lambda\,\dot\psi-\frac12\frac{\dot\psi^2}{\psi}+\frac13\,\psi\left[12\Lambda+b'\left(\psi\right)\right]=0,
\label{dotpsi}
\ee
respectively.
In the particular case where
\be
b\left(\psi\right)=-12\Lambda\,\psi\,,
\label{bchoice}
\ee
the potential becomes
\be
V\left(\varphi,\psi\right)=12\Lambda\left(\varphi-\psi\right)\,,
\label{potentialbchoice}
\ee
and Eq.~\eqref{dotpsi} can be divided by $\dot\psi$ and directly integrated over $t$ to obtain the solution
\be\label{solpsi}
\psi\left(t\right)=\psi_0 e^{-6 \sqrt\Lambda \, t} \left[e^{3\sqrt \Lambda \, \left(t-t_0\right)}-1\right]^2,
\ee
where $t_0$ and $\psi_0$ are constants of integration.
In the same case of Eq.~\eqref{bchoice}, where $b\left(\psi\right)=-12\Lambda\,\psi$, Eq.~\eqref{dotphi}
becomes
$\ddot\varphi+3\sqrt\Lambda\,\dot\varphi-4\Lambda\,\left(\varphi-\psi\right)=0$, and
upon using Eq.~\eqref{solpsi} it 
can be integrated to obtain the solution
\beq\label{solvarphiades1}
\varphi\left(t\right)=&&\psi_0 e^{-6  \sqrt\Lambda \,t_0}-\frac{2}{7} \psi_0 e^{-6 \sqrt\Lambda  t}\nonumber \\
&&-2 \psi_0 e^{-3 \sqrt\Lambda  \left(t+t_0\right)}+\varphi_0 e^{-4 \sqrt\Lambda  t}+\varphi_1 e^{\sqrt\Lambda  t},
\eeq
where $\varphi_0$ and $\varphi_1$ are constants of integration. The solutions for the scalar fields $\psi\left(t\right)$ and $\varphi\left(t\right)$ are plotted in Figs. \ref{fig:psi1} and \ref{fig:phi1}, respectively.

The solution is complete since $k$,
$a, \rho, p, V, \varphi$, and $\psi$ are known.

%%%%%%%%%%%%%%%%%%%%%%%%%%%%%%%%%%%%%%%%%%%%%%%%%%%%%%%%%%%%%%%%%%%%
\subsubsection{Solution with a simplified potential equation I}
\label{method2}
%%%%%%%%%%%%%%%%%%%%%%%%%%%%%%%%%%%%%%%%%%%%%%%%%%%%%%%%%%%%%%%%%%%%

\begin{figure}[h!]
\centering \includegraphics[scale=0.65]{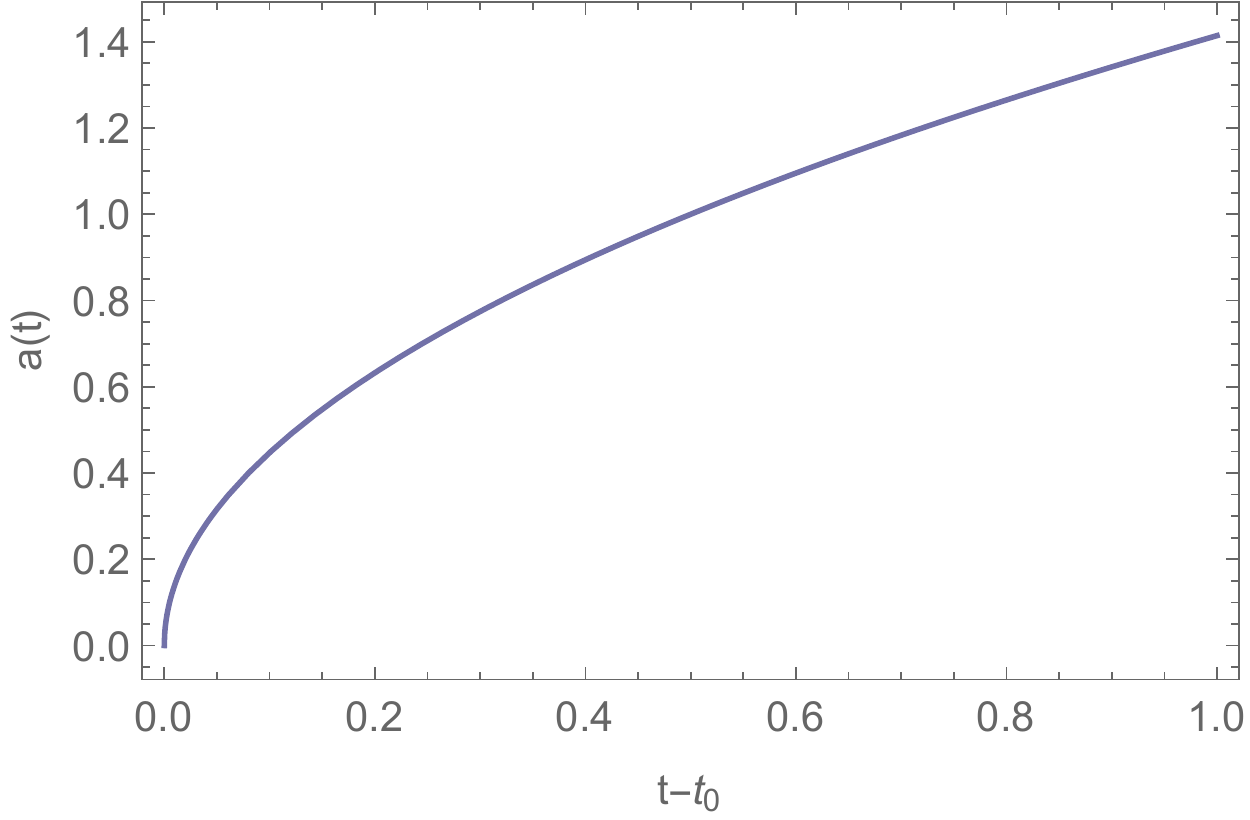}
\caption{Plot of the scale factor $a(t)$ for the case of the simplified potential I, given in Eq.~\eqref{a1pot},
for the values $a_0=1$, and $t_0=0$.}
\label{fig:a2}
\end{figure}

In this section, we set again the matter component to be vacuum, i.e., $\rho=p=0$, and we choose again a flat universe, $k=0$. Inspecting Eq.~\eqref{modF1}, we can see that it simplifies if the first two terms within the brackets cancel each other, i.e.,
\be
\dot H+2H^2=0,
\label{eqh1}
\ee
where we have used the definition for the Hubble function $H$, Eq.~\eqref{hubblep}.
Equation~\eqref{eqh1}
has the following solution for $H$,
\be
H=\frac{1}{2\left(t-t_0\right)},
\label{hs2}
\ee
where $t_0$ is an integration constant. Now, using again Eq.~\eqref{hubblep},
then Eq.~\eqref{hs2} can be integrated to obtain
\be
a\left(t\right)=a_0\sqrt{2\left(t-t_0\right)},
\label{a1pot}
\ee
where $a_0$ is an integration constant. This solution is plotted in Fig.\ref{fig:a2}.
 
The interest in this solution resides in the fact that, for a universe populated by radiation, we expect that the behavior of the scale factor is proportional to $\sqrt{t}$, but in this case the same behavior can be obtained with $\rho=0$.

\begin{figure}[h!]
\centering \includegraphics[scale=0.65]{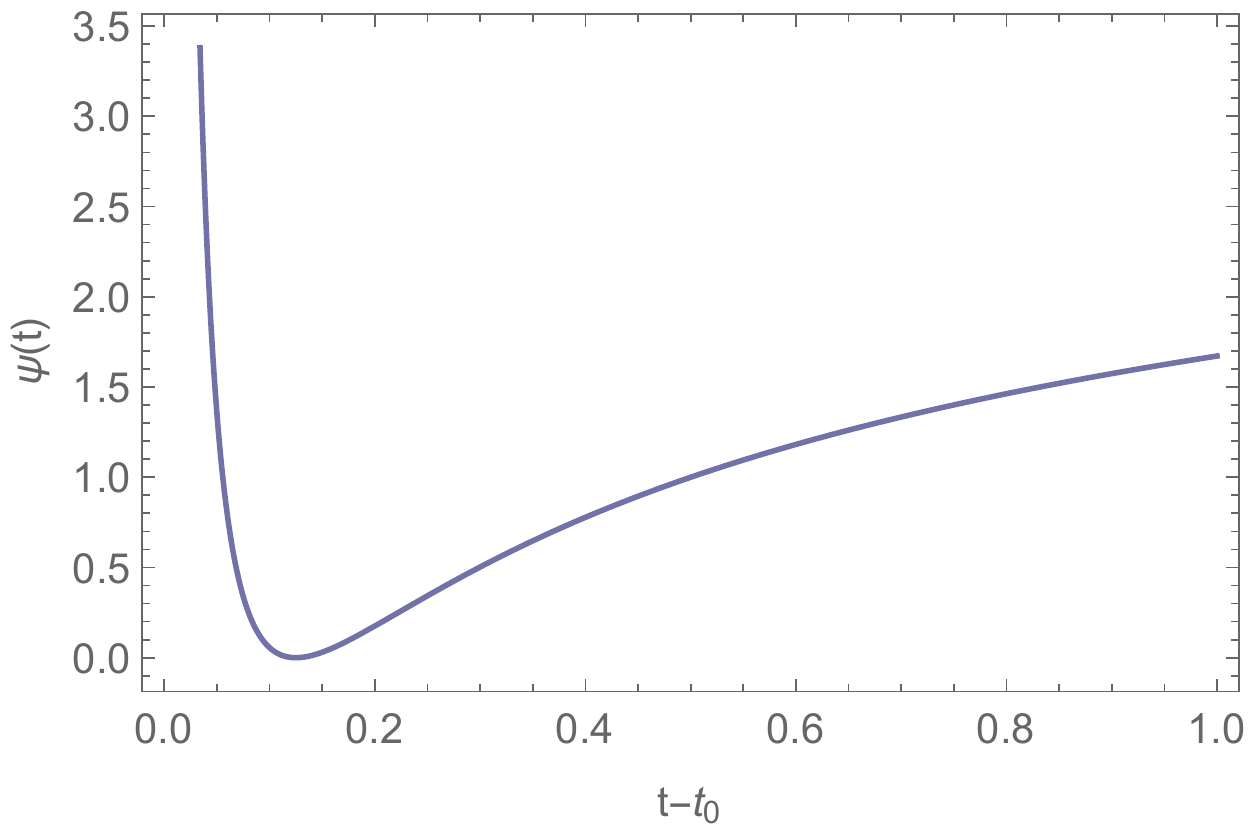}
\caption{Plot of the scalar field $\psi(t)$ for the case of the simplified potential I, given in Eq.~\eqref{psisimppot1},
for the values $\psi_0=\psi_1=1$, and $t_0=0$.}
\label{fig:psi2}
\end{figure}

\begin{figure}[h!]
\centering \includegraphics[scale=0.65]{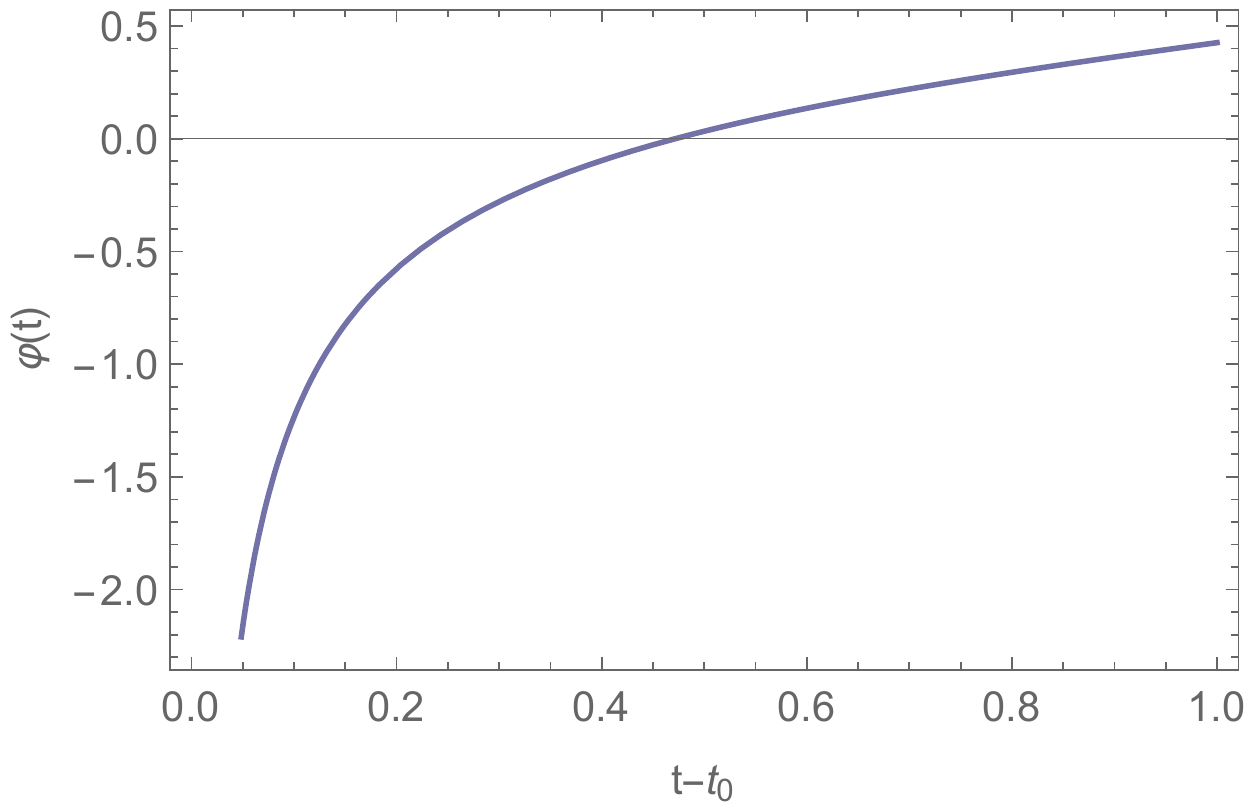}
\caption{Plot of the scalar field $\varphi(t)$ for the case of the simplified potential I, given in Eq.~\eqref{phisimppot1},
for the values $\varphi_0=\varphi_1=V_0=1$, and $t_0=0$.}
\label{fig:phi2}
\end{figure}

With this set of assumptions, the equation for the potential, Eq.~\eqref{modF1}, becomes 
\be\label{solV2}
V_\varphi=0,
\ee
which can be directly integrated to obtain
\be\label{solV13}
V\left(\varphi,\psi\right)=b\left(\psi\right),
\ee
where $b\left(\psi\right)$ is an arbitrary function on $\psi$. 
Using Eqs.~\eqref{a1pot} and~\eqref{solV13}, Eq.~\eqref{kgpsi0} for $\psi$ becomes
\be\label{ddotpsieq2}
\ddot\psi+\frac{3\dot\psi}{2\left(t-t_0\right)}-\frac{\dot\psi^2}{2\psi}+\frac{\psi}{3}b'\left(\psi\right)=0.
\ee
In the particular case where
\be
b\left(\psi\right)=V_0,
\ee
with $V_0$ a constant, the last term on the left-hand side of Eq.~\eqref{ddotpsieq2} vanishes and the equation can be integrated directly after dividing through by $\dot\psi$, yielding the solution
\be\label{psisimppot1}
\psi\left(t\right)=\frac{\psi_1 \left(\psi_0 \sqrt{2\left(t-t_0\right)}-1\right)^2}{2\left(t-t_0\right)},
\ee
where $\psi_0$ and $\psi_1$ are constants of integration. On the other hand, the equation for $\varphi$, namely Eq.~\eqref{kgphi0}, takes the form
\be\label{varphiother1}
\ddot\varphi+\frac{3\dot\varphi}{2\left(t-t_0\right)}-\frac{1}{3}\left[2b\left(\psi\right)-\psi b'\left(\psi\right)\right]=0.
\ee
Under the choice given in Eq.~\eqref{solV13} for $b\left(\psi\right)$, Eq.~\eqref{varphiother1}
decouples completely from the equation for $\psi$, yielding 
\be
\ddot\varphi+\frac{3\dot\varphi}{2\left(t-t_0\right)}-\frac{2}{3}V_0=0,
\ee
which can be directly integrated to obtain the following solution,
\be\label{phisimppot1}
\varphi\left(t\right)=\frac{2}{15}V_0t\left(t-2t_0\right)-\frac{\varphi_0}{\sqrt{2\left(t-t_0\right)}}+\varphi_1,
\ee
where $\varphi_0$ and $\varphi_1$ are constants of integration. The solutions for the scalar field $\psi\left(t\right)$ and $\varphi\left(t\right)$ are plotted in Figs. \ref{fig:psi2} and \ref{fig:phi2}, respectively. 

The solution is complete since 
$k,a, \rho, p, V, \varphi$, and $\psi$ are known.

%%%%%%%%%%%%%%%%%%%%%%%%%%%%%%%%%%%%%%%%%%%%%%%%%%%%%%%%%%%%%%%%%%%%
\subsubsection{Solution with a simplified potential equation II}
\label{method3}
%%%%%%%%%%%%%%%%%%%%%%%%%%%%%%%%%%%%%%%%%%%%%%%%%%%%%%%%%%%%%%%%%%%%

\begin{figure}[h!]
\centering \includegraphics[scale=0.65]{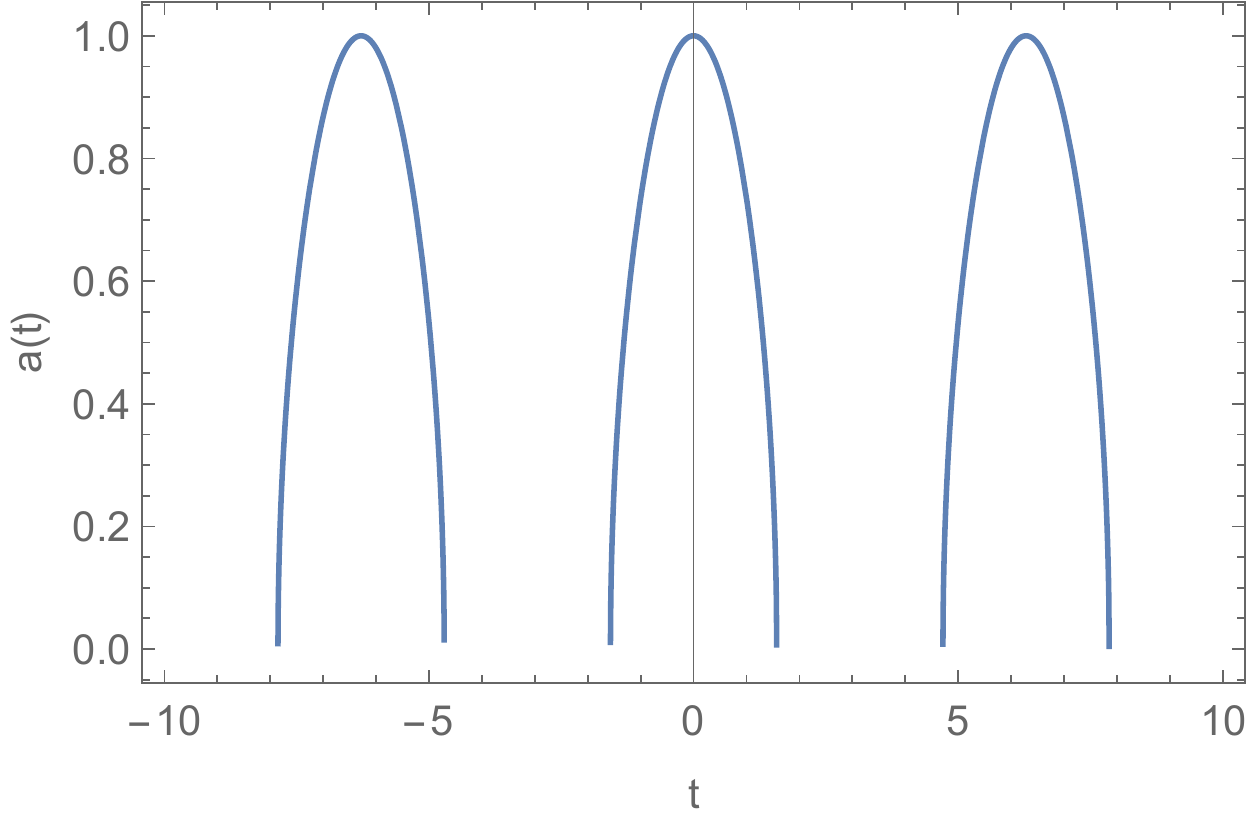}
\caption{Plot of the scale factor $a(t)$ for the case
 with a simplified potential equation II, given in Eq.~\eqref{scale3},
for the values $\Omega=1$,  $a_0=1$, and $t_0=0$.}
\label{fig:am3}
\end{figure}

In this section, we set again the matter component to be vacuum,
$\rho=p=0$, and we choose again a flat universe, $k=0$.
Taking into account Eq.~\eqref{modF1}, we look for solutions in which the first two
terms in the brackets are equal to a constant $\Omega^2/2$ say,
where the 2 enters for convenience. Then Eq.~\eqref{modF1}
is now
\be
\dot H+2H^2+\frac{\Omega^2}{2}=0\,,
\label{assumption2}
\ee
where we have used the definition for the Hubble function
$H$, Eq.~\eqref{hubblep}.
This condition yields the solution
\be
H=-{\frac{\Omega}{2}}\tan\left[\Omega\left(t-t_0\right)\right]\,,
\label{hnew2}
\ee
where $t_0$ is a constant of integration.
Now, using again Eq.~\eqref{hubblep},
then Eq.~\eqref{hnew2} can be integrated to obtain
the  scale factor,
\be\label{scale3}
a\left(t\right)=a_0\sqrt{\cos\left[\Omega\left(t-t_0\right)\right]},
\ee
where $a_0$ is a positive constant of integration, $a_0>0$.
The behavior of $a(t)$ in Eq.~\eqref{scale3} is given in 
Fig.~\ref{fig:am3}.
This solution is valid in between  times 
$t=-\frac{\pi}{2\Omega}+t_0$ and $t=\frac{\pi}{2\Omega}+t_0$,
or times that are translated from these ones
by $2\pi n$ with integer $n$. 
In between  times $t=\frac{\pi}{2\Omega}+t_0$ and $t=\frac{3\pi}{2\Omega}+t_0$,
or times that are translated from these ones
by $2\pi n$,
there are no physical solutions, since the scale factor in 
Eq.~\eqref{scale3} gets imaginary values.
The solution thus
represents a universe which starts expanding at $t=-\frac{\pi}{2\Omega}+t_0$, attains a maximum value at $t=t_0$, and then collapses again at $t=\frac{\pi}{2\Omega}+t_0$.
%, only to reappear back again at time 
%$t=\frac{3\pi}{2\Omega}+t_0$ when the process starts all over again
The importance of this solution for $a\left(t\right)$ stands on the fact that in the later times of the universe expansion, where we might approximate the distribution of matter to be vacuum and the geometry to be flat, it is possible to obtain solutions for the scale factor with a dependence on time different than the usual $t^{1/2}$ power-law expected in standard general relativity.

\begin{figure}[h!]
\centering \includegraphics[scale=0.65]{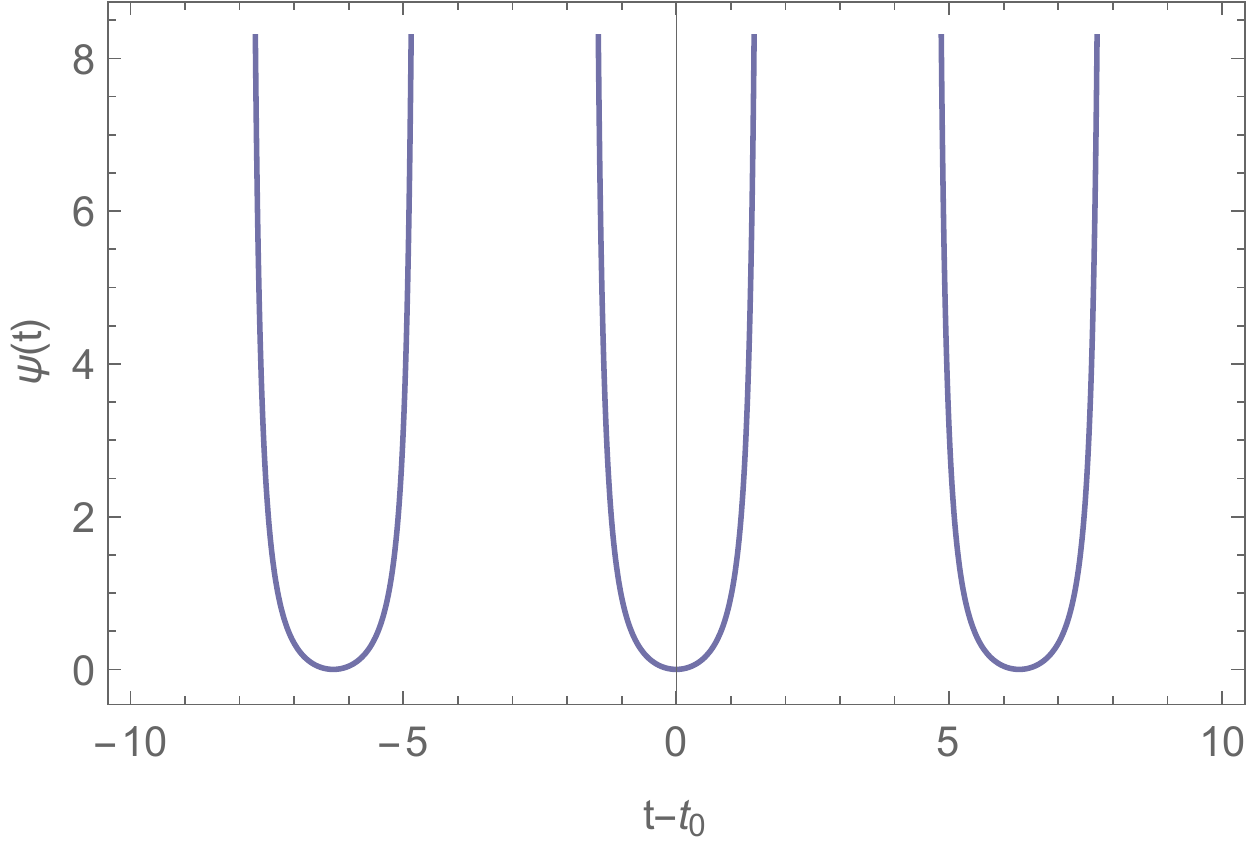}
\caption{Plot of the function $\psi(t)$  for the case
 with a simplified potential equation II, given in Eq.~\eqref{solpsiassump3}, for $\Omega=1$,
$\psi_0=0$, $\psi_1=1$, and
$t_0=0$.}
\label{fnumpsi}
\end{figure}

Assumption~\eqref{assumption2} leaves the equation for the potential, Eq.~\eqref{modF1}, in the form
\be
V_\varphi=-3\Omega^2,
\ee
which can be integrated to obtain
\be
V\left(\varphi,\psi\right)=-3\Omega^2\varphi+b\left(\psi\right),
\ee
where $b\left(\psi\right)$ is an arbitrary function of $\psi$. With these results, the equation for $\psi$, Eq.~\eqref{kgpsi0}, can be written as
\be
\ddot\psi-\frac32\Omega \tan\left[\Omega\left(t-t_0\right)\right]\dot\psi-\frac{\dot\psi^2}{2\psi}+\frac{\psi}{3}\left[b'\left(\psi\right)-3\Omega^2\right]=0.
\label{psiassump3}
\ee
In the particular case where
\be
b\left(\psi\right)=3\Omega^2\psi\,,
\label{b2special}
\ee
the potential becomes
\be
V\left(\varphi,\psi\right)=-3\Omega^2\left(\varphi-\psi\right)\,,
\label{potiispec}
\ee
the derivatives are $V_\varphi=-V_\psi$ and the last term in the
left-hand side of Eq.~\eqref{psiassump3}
vanishes. The equation can then be integrated analytically to obtain a solution for $\psi$ of the form
\beq\label{solpsiassump3}
\psi\left(t\right)=&&\psi_1\sec\left[\Omega\left(t-t_0\right)\right]\left\{\sqrt{\cos\left[\Omega\left(t-t_0\right)\right]}\times\right.\nonumber \\
&&\times\left[\psi_0\frac{\Omega}{\sqrt2}-\sqrt{2}E\left({\frac{\Omega}{2}}\left(t-t_0\right)|2\right)\right]\nonumber \\
&&\left.+\sqrt{2}\sin\left[\Omega\left(t-t_0\right)\right]\right\}^2,
\eeq
where $\psi_0$ and $\psi_1$ are constants of integration and $E\left(a|b\right)$ is the incomplete elliptic integral of the second kind defined as
$
E\left(a|b\right)=\int_0^a\sqrt{1-b\sin^2\theta}d\theta$.
Figure~\ref{fnumpsi} plots $\psi$ as a function of $t$.
As for the scale factor $a(t)$ 
the solution for $\psi(t)$ is valid in between  times 
$t=-\frac{\pi}{2\Omega}+t_0$ and $t=\frac{\pi}{2\Omega}+t_0$,
or times that are translated from these ones
by $2\pi n$ with integer $n$. 

\begin{figure}[h!]
\centering \includegraphics[scale=0.65]{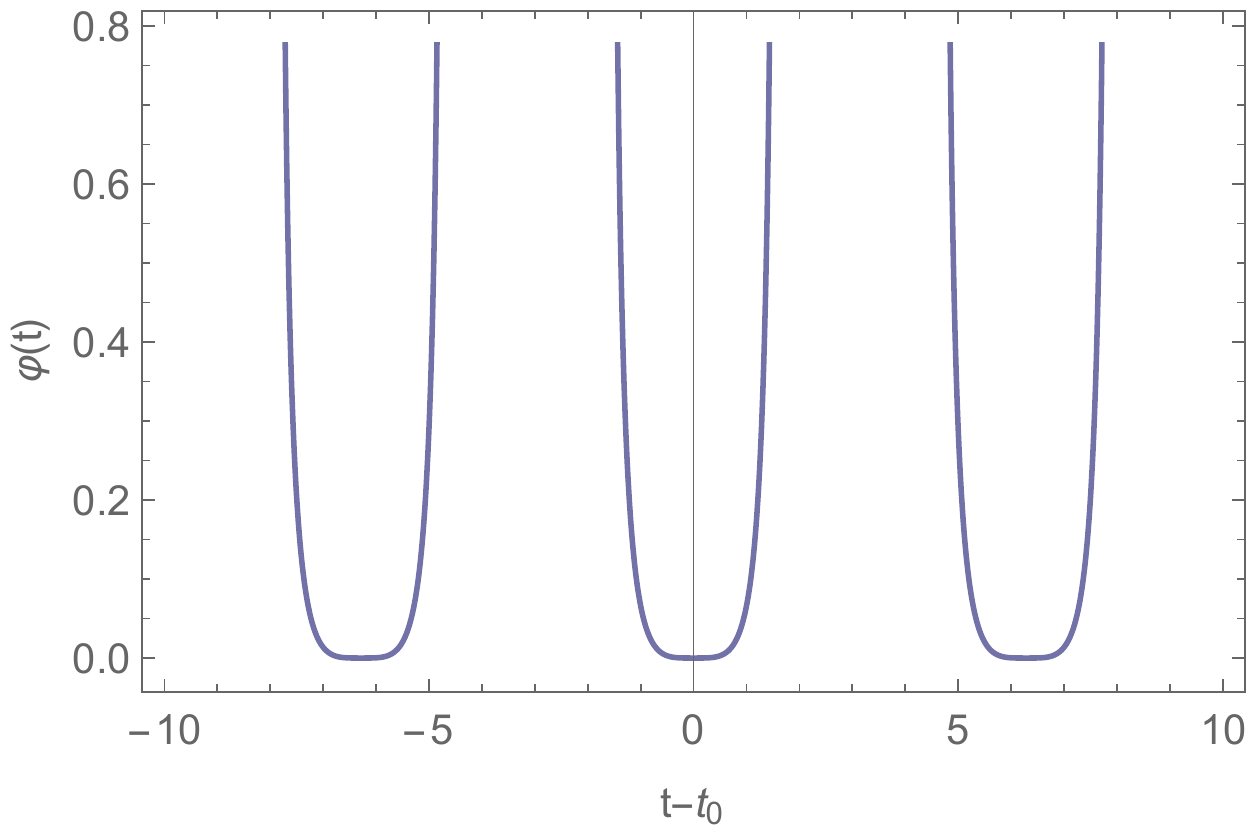}
\caption{Plot of the function $\varphi(t)$ for the case
 with a simplified potential equation II, from numerical results for the integration of Eq.~\eqref{numphieq2} for $\Omega=1$,
$\psi_0=0$, $\psi_1=1$, $\varphi_0=0$, $\dot{\varphi_0}=0$, and 
$t_0=0$.}
\label{fnumphi2}
\end{figure}

Finally, the equation for $\varphi$, Eq.~\eqref{kgphi0}, becomes
\be
\ddot\varphi-\frac32\Omega \tan\left[\Omega\left(t-t_0\right)\right]\dot\varphi+\frac{\psi}{3}b'\left(\psi\right)-\frac{2}{3}b\left(\psi\right)+\Omega^2\varphi=0.
\label{psieq3}
\ee
Choosing the same function $b\left(\psi\right)$, see Eq.~\eqref{b2special}, Eq.~\eqref{psieq3}
simplifies to
\be\label{numphieq2}
\ddot\varphi-\frac32\Omega \tan\left[\Omega\left(t-t_0\right)\right]\dot\varphi+\Omega^2\left(\varphi-\psi\right)=0.
\ee
This equation does not have an analytical solution. However, it can be integrated numerically. In Fig.~\ref{fnumphi2} we plot the  solution  for $\varphi(t)$ for a specific choice of the parameters
and with $\varphi(0)\equiv\varphi_0=0$ and  $\dot{\varphi}(0)\equiv\dot{\varphi_0}=0$.
%\footnote{In doing so we note that, being Eq.~\eqref{numphieq2} a
%second order ordinary linear differential equation, it might present
%complex solutions. By the superposition principle, however, we can
%always restrict ourselves to the particular solution associated to the
%real part of $\varphi$ without loss of generality ($\psi$ is defined
%real in the domain of integration). These solutions are the ones
%plotted in Fig.~\ref{fnumphi2}.}.
As for the scale factor $a(t)$ and for the scalar
$\psi(t)$ the solution for $\varphi(t)$ is valid in between  times 
$t=-\frac{\pi}{2\Omega}+t_0$ and $t=\frac{\pi}{2\Omega}+t_0$,
or times that are translated from these ones
by $2\pi n$ with integer $n$.

The solution is complete since 
$k,a, \rho, p, V, \varphi$, and $\psi$ are known.

%%%%%%%%%%%%%%%%%%%%%%%%%%%%%%%%%%%%%%%%%%%%%%%%%%%%%%%%%%%%%%%%%%%%
\subsubsection{Solution with simplified scalar field equations}
\label{method4}
%%%%%%%%%%%%%%%%%%%%%%%%%%%%%%%%%%%%%%%%%%%%%%%%%%%%%%%%%%%%%%%%%%%%

In this section, we set again the matter component to be vacuum,
$\rho=p=0$, and choose a flat universe, $k=0$.

\begin{figure}[h!]
\centering \includegraphics[scale=0.65]{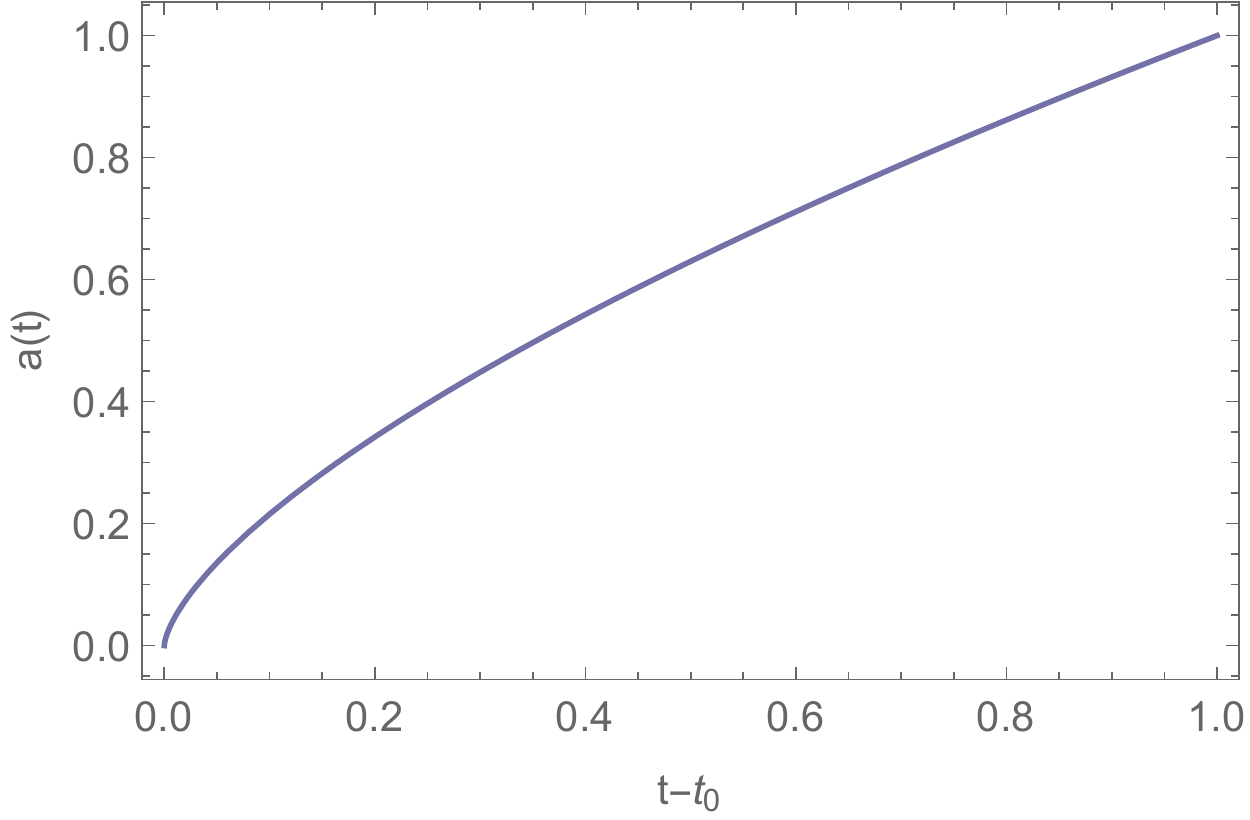}
\caption{Plot of the scale factor $a(t)$ for the case
 with simplified scalar field equations, given in Eq.~\eqref{scale31},
for the values $a_0=1$, and $t_0=0$.}
\label{fig:a4}
\end{figure}

For simplicity, we introduce a potential of the following form
\be\label{simpotential}
V=V_0\left(\varphi-\psi\right)^2,
\ee
where $V_0$  is a constant.
Using this potential, the scalar field Eqs.~\eqref{kgphi0} and~\eqref{kgpsi0},
with the notation for the Hubble function $H$ given in
Eq.~\eqref{hubblep},
simplify to
\be\label{simvarphi}
\ddot\varphi+3H\dot\varphi=0,
\ee
and
\be\label{simpsi}
\ddot\psi+3H\dot\psi-\frac{\dot\psi^2}{2\psi}=0,
\ee
respectively. The first of these equations, Eq.~\eqref{simvarphi}, can be integrated directly after dividing through by $\dot\varphi$ to obtain the equation
\be\label{soldphi1}
\dot\varphi=\frac{\varphi_0}{a^3},
\ee
where $\varphi_0$ is an arbitrary constant of integration. To integrate the second equation, Eq.~\eqref{simpsi}, we divide it through by $\dot\psi$ which results in
\be\label{simpsi2}
\frac{\dot\psi}{\sqrt{\psi}}=\frac{\psi_0}{a^3},
\ee
where $\psi_0$ is a constant of integration. Equating Eq.~\eqref{simpsi}
to Eq.~\eqref{soldphi1}, one obtains a relation between the scalar fields given by
\be\label{soldpsi1}
\frac{\dot\psi}{\psi_0\sqrt{\psi}}=\frac{\dot\varphi}{\varphi_0},
\ee
which can be integrated to yield $\varphi=2\frac{\varphi_0}{\psi_0}\sqrt{\psi}+\varphi_1$,
which upon inverting gives, 
\be\label{relsca1}
\psi=\left(\frac{\psi_0}{2\varphi_0}\right)^2\left(\varphi-\varphi_1\right)^2,
\ee
%\be\label{relsca1}
%\varphi=2\frac{\varphi_0}{\psi_0}\sqrt{\psi}+\varphi_1,
%\ee
where $\varphi_1$ is a constant of integration. 
Inserting the potential~\eqref{simpotential} in Eq.~\eqref{modF1}, we obtain
a relation between the Hubble function $H$, and so the scale factor,  and the scalar fields as
\be\label{fr1}
\dot H+2H^2=\frac{V_0}{3}\left(\varphi-\psi\right).
\ee
Inserting   Eq.~\eqref{relsca1} into Eq.~\eqref{fr1}
we obtain
\be\label{fr2}
\dot H+2H^2=\frac{V_0}{3}\left[\varphi-\left(\frac{\psi_0}{2\varphi_0}\right)^2\left(\varphi-\varphi_1\right)^2\right].
\ee
Now, after differentiating Eq.~\eqref{fr2}
with respect to time, using Eq.~\eqref{soldphi1} to cancel the $\dot\varphi$ term, and deriving again with respect to time, we can write Eq.~\eqref{fr2} as
\be\label{eqndddH}
\dddot H+7\ddot H H+4\dot H^2+12 \dot H H^2=-\frac{V_0\psi_0^2}{6a^6}.
\ee
This is a fourth degree nonlinear differential equation for $a$. It may not have an
analytical general solution, but one can try a particular solution through 
 an ansatz of the form $a=a_0t^\alpha$, for $a_0$ a constant and $\alpha$ a number.
This ansatz put into Eq.~\eqref{eqndddH} leads to
$\left(-12\alpha^3+18\alpha^2-6\alpha\right)\frac{1}{t^4}=-
\frac{V_0\psi_0^2}{6a_0^6}\frac{1}{t^{6\alpha}}$.
Equating the exponents gives $\alpha=2/3$.
So the scale factor $a(t)$ behaves as
\be\label{scale31}
a(t)=a_0\,t^{2/3}\,
\ee
This solution is plotted in Fig. \ref{fig:a4}.
Moreover, one finds that
$12\alpha^3-18\alpha^2+6\alpha=-\frac{4}{9}$, and so 
equating the factors of the  equation above yields 
a constraint for the constants appearing in Eq.~\eqref{eqndddH}, namely,
\be
\frac{V_0\psi_0^2}{6a_0^6}=-\frac{4}{9}\,.
\ee

\begin{figure}[h!]
\centering \includegraphics[scale=0.65]{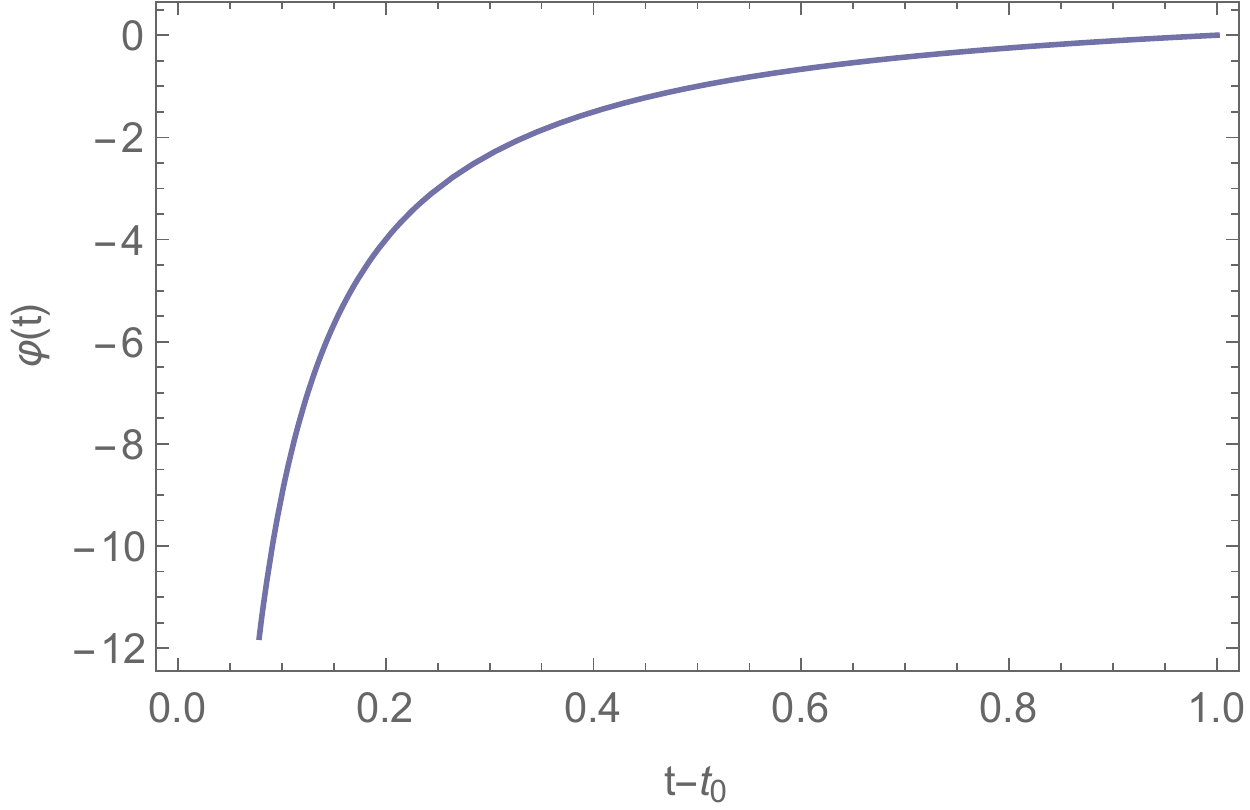}
\caption{Plot of the scalar field $\varphi(t)$ for the case
 with simplified scalar field equations, given in Eq.~\eqref{invphi4},
for the values $\varphi_0=\varphi_1=1$, and $t_0=0$.}
\label{fig:phi4}
\end{figure}

\begin{figure}[h!]
\centering \includegraphics[scale=0.65]{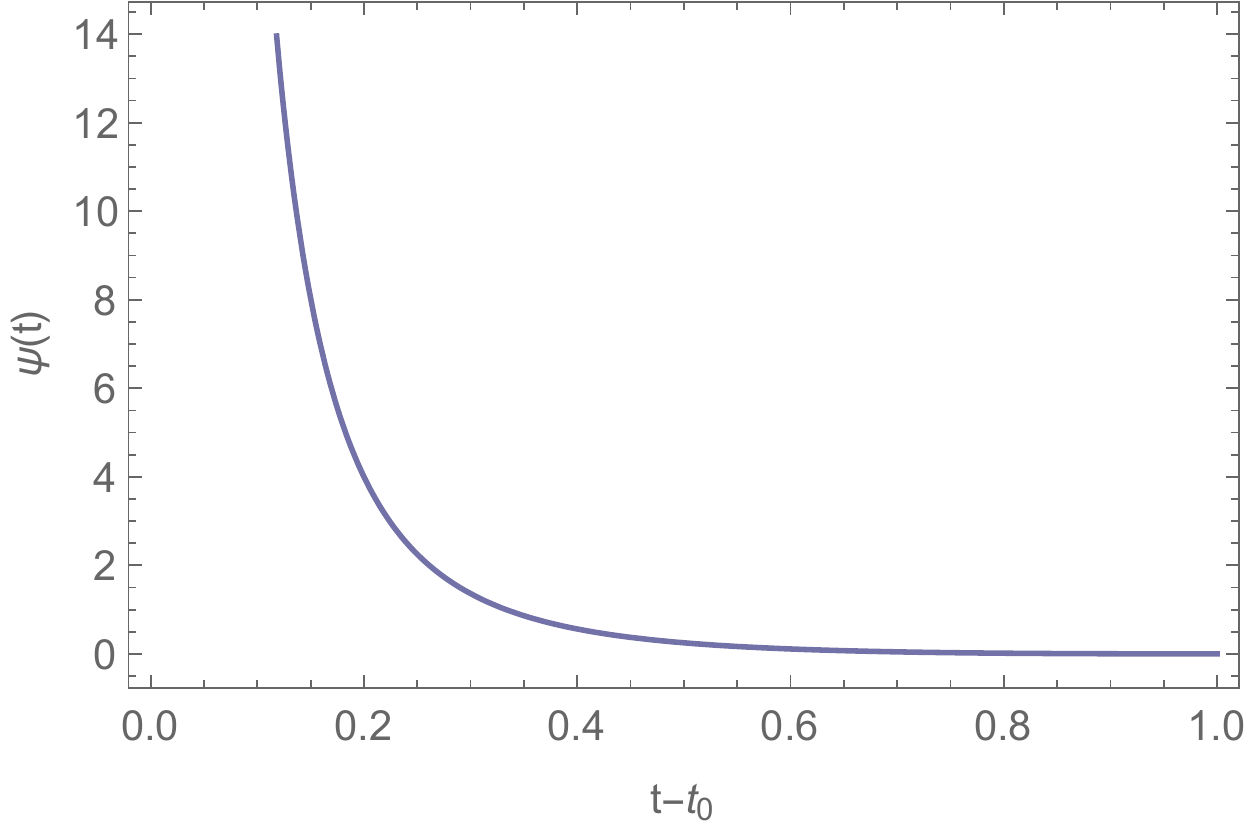}
\caption{Plot of the scalar field $\psi(t)$ for the case
 with simplified scalar field equations, given in Eq.~\eqref{invpsi4},
for the values $\psi_0=\psi_1=1$, and $t_0=0$.}
\label{fig:psi4}
\end{figure}

The solution for the scale factor,  Eq.~\eqref{scale3},
allows us to integrate Eqs.~\eqref{soldphi1} and~\eqref{soldpsi1} directly over time, leading to the following solutions for the scalar fields
\be\label{invphi4}
\varphi\left(t\right)=-\frac{\varphi_0}{a_0^3t}+\varphi_1 ,
\ee
and
\be\label{invpsi4}
\psi\left(t\right)=\left(-\frac{\psi_0}{2a_0^3t}+\frac{\psi_1}{2}\right)^2,
\ee
respectively.These solutions are plotted in Figs. \ref{fig:phi4} and \ref{fig:psi4}, respectively.
This solution is interesting because, for a universe dominated by
matter, i.e., $p=0$, one expects the scale factor to behave as
$t^\frac{2}{3}$, in this case we obtain this behavior considering
$\rho=0$.

The solution is complete since 
$k,a, \rho, p, V, \varphi$, and $\psi$ are known.

%%%%%%%%%%%%%%%%%%%%%%%%%%%%%%%%%%%%%%%%%%%%%%%%%%%%%%%%%%%%%%%%%%%%
\subsubsection{Solution with nonflat geometry}
\label{method5}
%%%%%%%%%%%%%%%%%%%%%%%%%%%%%%%%%%%%%%%%%%%%%%%%%%%%%%%%%%%%%%%%%%%%

\begin{figure}[h!]
\centering \includegraphics[scale=0.75]{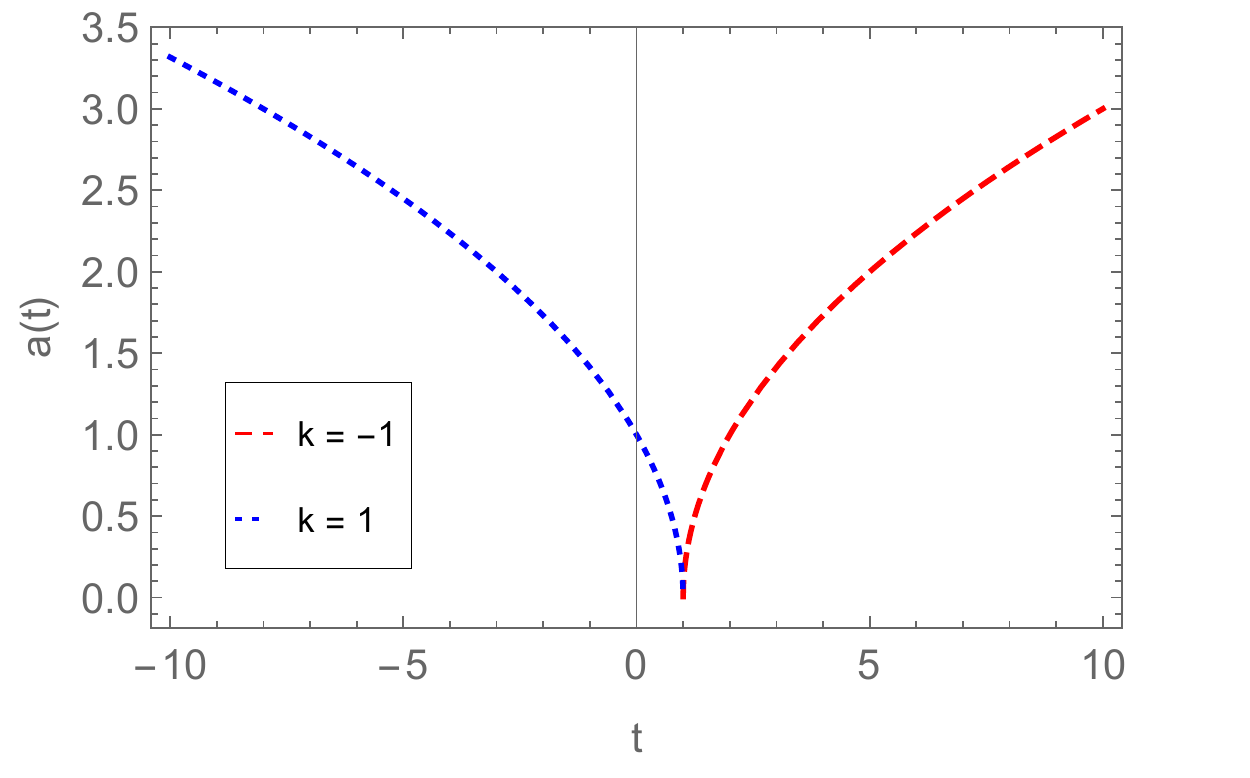}
\caption{Plot of the scale factor~\eqref{scale5} for $a_0=1$, $t_0=0$.}
\label{fig:a5}
\end{figure}

In this section, we
set again the matter component to be vacuum, $\rho=p=0$. Now, we consider a nonflat geometry, i.e.,
\be\label{newk5}
k=1,-1\,.
\ee
Looking at Eq.~\eqref{modF1}, we can try to find a simplified equation for the potential considering that the first three terms inside the brackets cancel each other. Using
Eq.~\eqref{hubblep} this ansatz can be written as
\be\label{newhdot1}
\dot H+2H^2+\frac{k}{a^2}=0\,.
\ee
Equation~\eqref{newhdot1} is in fact a nonlinear ordinary differential equation of the second order for $a\left(t\right)$. Integrating this equation, we find the solution
\be\label{scale5}
a\left(t\right)=\sqrt{\frac{a_0^2-k^2\left(t-t_0\right)}{k}},
\ee
where $k$ is one of the two values given in
Eq.~\eqref{newk5}, $a_0$ and $t_0$ are constants of integration, and we have
not considered the solution with negative sign for the scale factor $a(t)$. Notice that for $k=1$, there is an end value of $t$, given by $t_e=a_0^2+t_0$, for which the scale factor drops to zero, ending the solution,
as for $t>t_e$ it becomes a pure imaginary number. The same happens for $k=-1$ when $t<t_e$. Therefore, the solution for $k=1$ is only defined for $-\infty<t<t_e$ and the solution for $k=-1$ is only defined for $t_e<t<+\infty$, see Fig.~\ref{fig:a5}.
We know from observations that $k\sim 0$ is a very good approximation. However, the field equations \eqref{genein2} depend on $k$ in such a way that if $k$ varies even slightly from $0$, the structure of the equations changes dramatically. The relevance of the solution we have found, therefore, is to provide a glimpse of the role of spatial curvature in these theories.

Taking into account Eq.~\eqref{newhdot1}, the equation for the potential, Eq.~\eqref{modF1}, reduces to $V_\varphi=0$. This can be directly integrated over $\varphi$ to obtain
\be
V\left(\varphi,\psi\right)=b\left(\psi\right),
\ee
where $b\left(\psi\right)$ is an arbitrary function of $\psi$.
With these results, %the equation for $\psi$ given by the evolution equation, Eq.~\eqref{kgpsi0}, becomes
%$\ddot\psi+\frac{2k^2\left(t-t_0\right)}{k^2\left(t-t_0\right)^2-a_0^2}\dot\psi-\frac{\dot\psi^2}{2\psi}+\frac{\psi}{3}b'\left(\psi\right)=0$,
%and the 
% equation for $\varphi$, Eq.~\eqref{kgphi0}, takes the form
%$\ddot \varphi-\frac{3k^2\left(t-t_0\right)}{a_0^2-k^2\left(t-t_0\right)^2}\dot\varphi+\frac{\psi}{3}b'\left(\psi\right)-\frac{2}{3}b\left(\psi\right)=0$.
%To make progress
%in solving these differential equations for $\psi$ and $\varphi$ 
we further assume $b'\left(\psi\right)=0$, i.e.,
$b\left(\psi\right)=V_0$, so that 
\be
V\left(\varphi,\psi\right)=V_0,
\ee
with $V_0$ is a constant,

Then, Eq.~\eqref{kgpsi0}  becomes
\be
\ddot\psi+\frac{2k^2\left(t-t_0\right)}{k^2\left(t-t_0\right)^2-a_0^2}\dot\psi-\frac{\dot\psi^2}{2\psi}=0.
\ee
This equation can be solved analytically and we obtain
\beq\label{psi5}
\psi\left(t\right)=&&\frac{\psi_1}{a_0^2-k^2\left(t-t_0\right)^2} \left[\left(a_0^4 \psi_0 k^2+1\right) \left(t-t_0\right)^2\right.\nonumber \\
&&-\left.a_0^6 \psi_0\right] e^{ 2 \tanh^{-1}\left[\frac{t-t_0}{a_0^2 \sqrt{\psi_0} \sqrt{a_0^2-k^2 \left(t-t_0\right)^2}}\right]},
\eeq
where $\psi_0$ and $\psi_1$ are constants of integration.

\begin{figure}[h!]
\centering \includegraphics[scale=0.65]{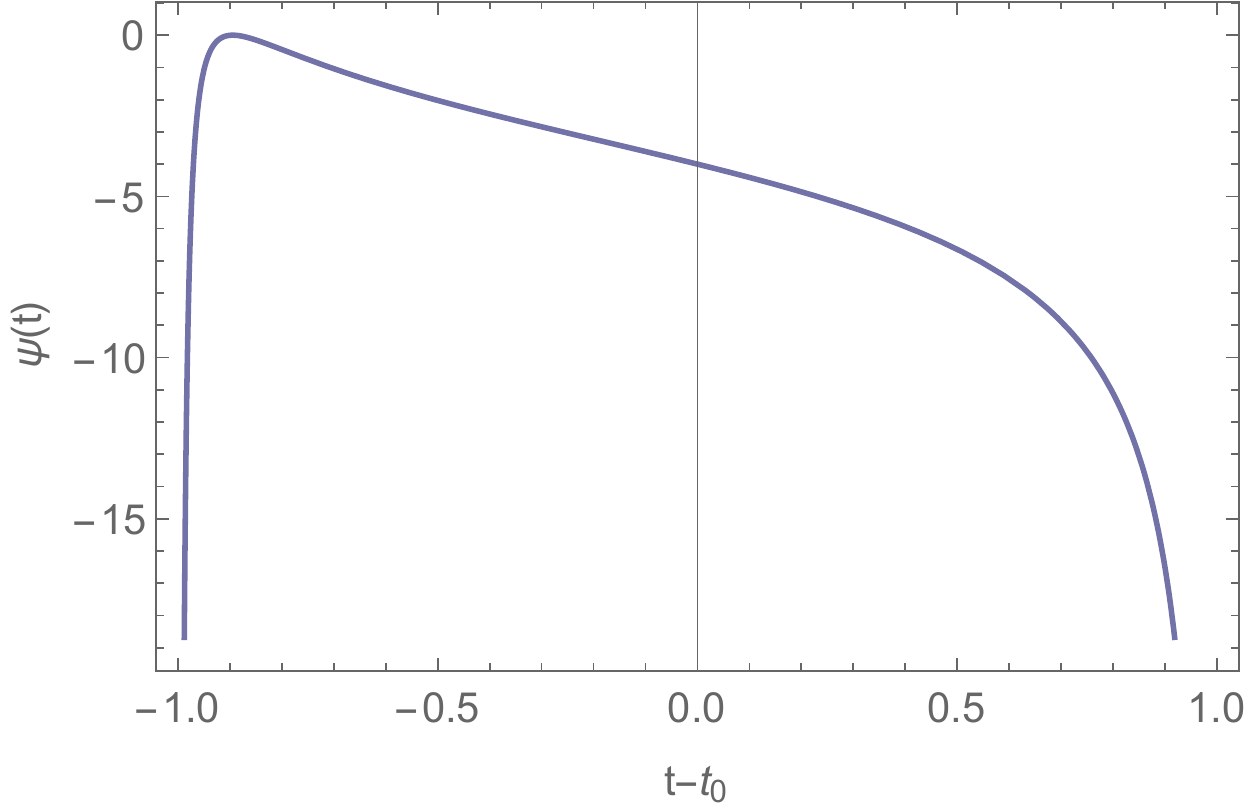}
\caption{Plot of the scalar field $\psi(t)$ for the case
 with nonflat geometry, given in Eq.~\eqref{psi5},
for the values $\psi_0=\psi_1=a_0=1$, and $t_0=0$.}
\label{fig:psi5}
\end{figure}

\begin{figure}[h!]
\centering \includegraphics[scale=0.65]{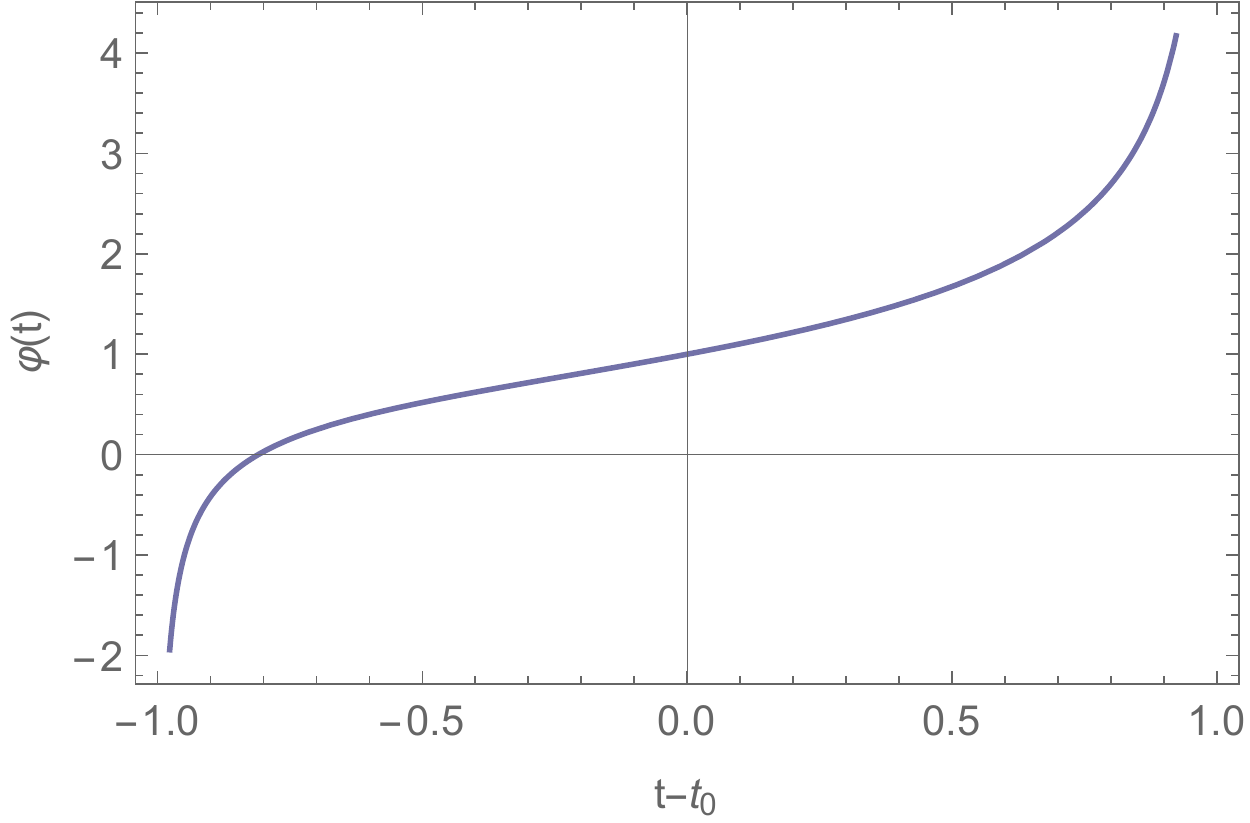}
\caption{Plot of the scalar field $\varphi(t)$ for the case
 with nonflat geometry, given in Eq.~\eqref{varphi5},
for the values $\varphi_0=\varphi_1=a_0=V_0=1$, and $t_0=0$.}
\label{fig:phi5}
\end{figure}

On the other hand, Eq.~\eqref{kgphi0}, simplifies to
\be\label{numphieq3}
\ddot \varphi-\frac{3k^2\left(t-t_0\right)}{a_0^2-k^2\left(t-t_0\right)^2}\dot\varphi
-\frac{2}{3}V_0=0.
\ee
This equation is once again decoupled from $\psi$ and can be integrated directly to obtain
\beq\label{varphi5}
\varphi\left(t\right)=&&\varphi_1+\frac{ t-t_0}{12a_0^2k \sqrt{a_0^2-k^2 \left(t-t_0\right)^2}}
\times\nonumber \\
&&\left[3 a_0^4 V_0 \tan ^{-1}\left(\frac{k\left(t-t_0\right)}{\sqrt{a_0^2-k^2 \left(t-t_0\right)^2}}\right)\right.\\
&&\left.+k \left(a_0^2 V_0 \left(t-t_0\right) \sqrt{a_0^2-k^2 \left(t-t_0\right)^2}+\varphi_0\right)\right],
\nonumber
\eeq
where $\varphi_0$ and $\varphi_1$ are constants of integration. The solutions for $\psi\left(t\right)$ and $\varphi\left(t\right)$ are plotted in Figs. \ref{fig:psi5} and \ref{fig:phi5}, respectively.

The solution is complete since 
$k,a, \rho, p, V, \varphi$, and $\psi$ are known.

%%%%%%%%%%%%%%%%%%%%%%%%%%%%%%%%%%%%%%%%%%%%%%%%%%%%%%%%%%%%%%%%%%%%
\subsubsection{Solution with perfect fluid matter}
\label{method6}
%%%%%%%%%%%%%%%%%%%%%%%%%%%%%%%%%%%%%%%%%%%%%%%%%%%%%%%%%%%%%%%%%%%%

In this section we assume the presence of a perfect fluid, Eq.(\ref{perffluid}), with an equation
of state as given in Eq.~\eqref{eos}. Then
Eq.~\eqref{econs} provides solutions for $\rho$ and $p$ as
\be\label{rho6}
\rho=\rho_0a^{-3\left(1+w\right)}\,,
\ee
\be\label{p6}
p=p_0a^{-3\left(1+w\right)},
\ee
respectively,
where $\rho_0$ and $p_0$ are constants of integration that are related to each other by the equation of state~\eqref{eos}.
We also assume a flat universe, $k=0$.

Now, using Eqs.~\eqref{rho6} and~\eqref{p6} we have that the energy-momentum tensor
$T_{ab}$ has the  trace $T$ given by
$T=\left(-\rho+3p\right)=-\rho_0\left(1-3w\right)a^{-3\left(w+1\right)}$.
Then Eq.~\eqref{kgphi0} becomes
\be\label{newvarphi6}
\ddot\varphi+3H\dot\varphi-\frac{1}{3}\left[2V-\psi V_\psi-\varphi V_\varphi\right]=\frac{\kappa^2\rho_0}{3}\left(1-3w\right)a^{-3\left(w+1\right)}.
\ee
In addition, Eq.~\eqref{kgpsi0} for $\psi$ and Eq.~\eqref{modF1} for $V_\varphi$ remain the same.
So, we just have to solve Eq.~\eqref{newvarphi6} for any of the particular cases discussed before,
the other functions remaining the same. We consider two cases, namely,
the de Sitter expansion of Sec.~\ref{method1} and
the simplified equation for the potential I of Sec.~\ref{method2}.

\begin{figure}[h!]
\centering \includegraphics[scale=0.65]{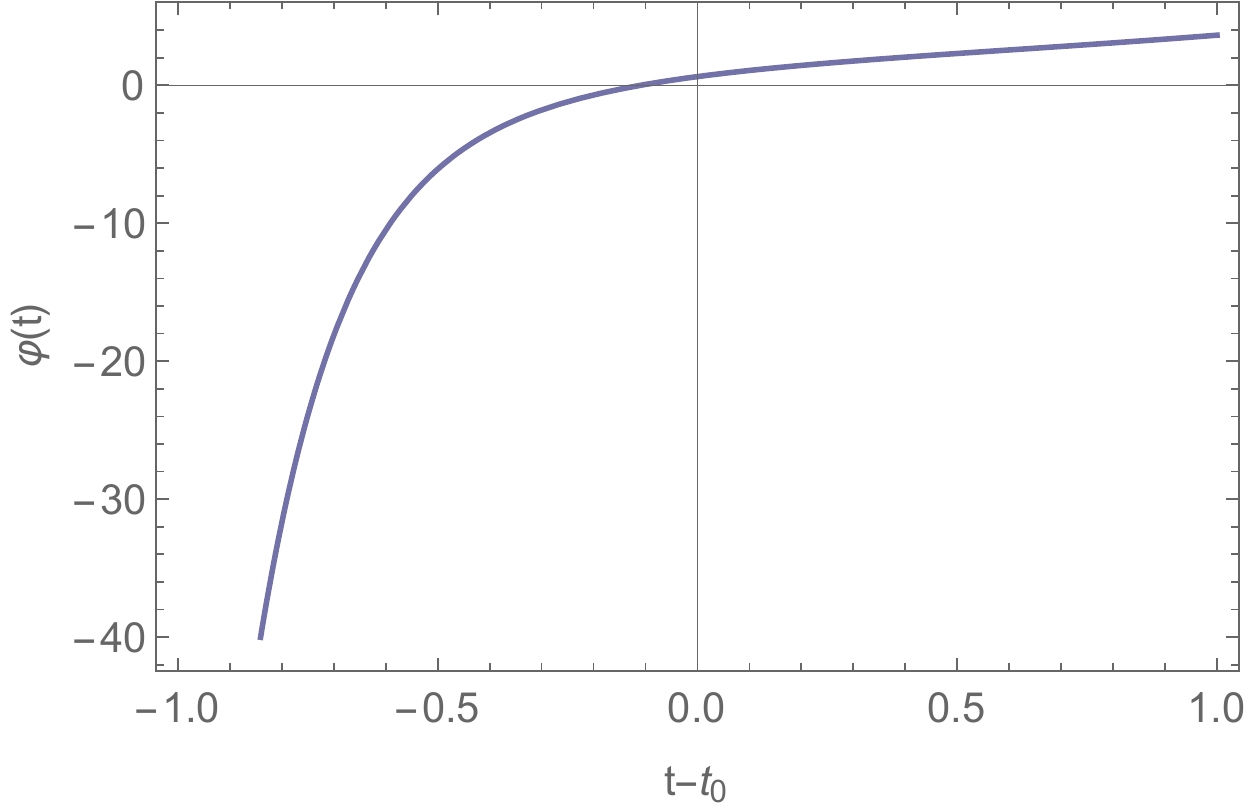}
\caption{Plot of the scalar field $\varphi(t)$ for the de Sitter case with perfect fluid matter, given in Eq.~\eqref{solvarphiades11},
for the values $\kappa=\rho_0=a_0=\Lambda=\psi_0=\varphi_0=\varphi_1=1$, and $t_0=w=0$.}
\label{fig:phimat1}
\end{figure}

If we assume a de Sitter expansion, see Sec.~\ref{method1}, then the results
from Eq.~\eqref{scalefactordes1} to~\eqref{solpsi} hold,
in particular $a(t)$ behaves as in Fig.~\ref{fig:a1} and
$\varphi(t)$ as in Fig.~\ref{fig:phi1}. 
However, instead of Eq.~\eqref{solvarphiades1} and the corresponding Fig.~\ref{fig:psi1},
we have to solve 
Eq.~\eqref{newvarphi6}, which
can be integrated analytically to yield
\beq\label{solvarphiades11}
\varphi\left(t\right)=&&-\frac{\kappa^2\rho_0\left(a_0e^{\sqrt{\Lambda} t}\right)^{-3w}e^{-3\sqrt{\Lambda} t}}{3a_0^3{\Lambda}\left(4+3w\right)}
    \nonumber\\
&&+\psi_0e^{-6\sqrt{\Lambda} t_0}-\frac{2}{7}\psi_0e^{-6\sqrt{\Lambda} t}
    \\
&&-2\psi_0e^{-3\sqrt{\Lambda}\left(t+t_0\right)}+\varphi_0e^{-4\sqrt{\Lambda} t}+\varphi_1e^{\sqrt{\Lambda} t}.\nonumber 
\eeq
This solution is plotted in Fig.~\ref{fig:phimat1}.

\begin{figure}[h!]
\centering \includegraphics[scale=0.65]{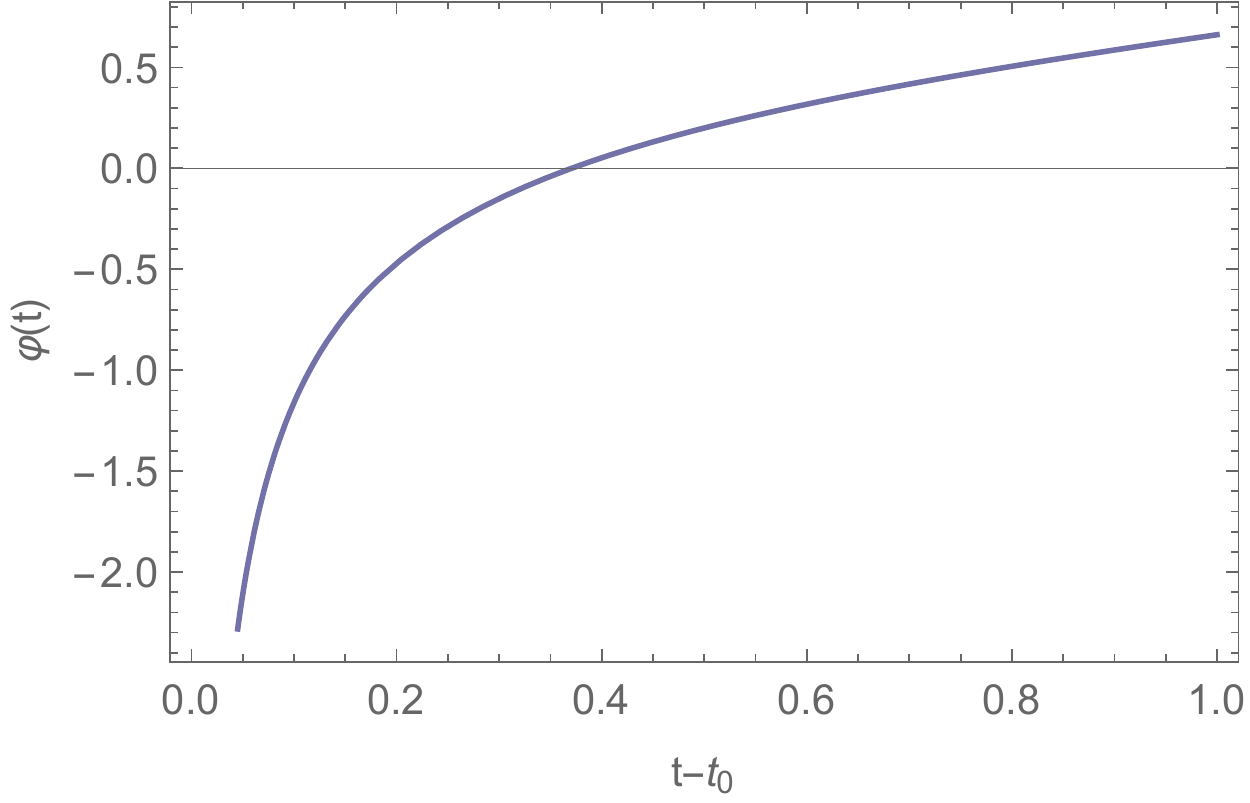}
\caption{Plot of the scalar field $\varphi(t)$ for the case
 with a simplified potential equation with perfect fluid matter, given in Eq.~\eqref{x6},
for the values  $\kappa=\rho_0=a_0=V_0=\varphi_0=\varphi_1=1$, and $t_0=w=0$.}
\label{fig:phimat2}
\end{figure}

If we assume a simplified equation for the potential I, see Sec.~\ref{method2}, then the results
from Eq.~\eqref{eqh1} to~\eqref{psisimppot1} also hold,
in particular $a(t)$ behaves as in Fig.~\ref{fig:a2} and
$\varphi(t)$ as in Fig.~\ref{fig:phi2}. 
Again, instead of Eq.~\eqref{varphiother1}  and the corresponding Fig.~\ref{fig:psi2}, we have to solve 
Eq.~\eqref{newvarphi6}, which
can be integrated analytically to yield
\beq\label{x6}
\varphi\left(t\right)=\frac{2}{15} V_0 \left(t-t_0\right)^2-\frac{\varphi_0}{\sqrt{2\left( t-t_0\right)}}+\varphi_1
\nonumber \\
-\frac{2\kappa ^2 \rho_0 \left(t-t_0\right) }{3 a_0^2 (3 w-2)}\; \left(a_0 \sqrt{2\left( t-t_0\right)}\right)^{-\left(3w+1\right)}\,.
\eeq
This solution is plotted in Fig.~\ref{fig:phimat2}.

The solutions are complete since 
$k,a, \rho, p, V, \varphi$, and $\psi$ are known.

%%%%%%%%%%%%%%%%%%%%%%%%%%%%%%%%%%%%%%%%%%%%%%%%%%%%%%%%%%%%%%%%%%%%
\section{Pulling the previous solutions into generalized hybrid gravity and the form of $f\left(R,\mathcal{R}\right)$}\label{secIV}
%%%%%%%%%%%%%%%%%%%%%%%%%%%%%%%%%%%%%%%%%%%%%%%%%%%%%%%%%%%%%%%%%%%%

%%%%%%%%%%%%%%%%%%%%%%%%%%%%%%%%%%%%%%%%%%%%%%%%%%%%%%%%%%%%%%%%%%%%
\subsection{The main equation for generalized hybrid gravity}
%%%%%%%%%%%%%%%%%%%%%%%%%%%%%%%%%%%%%%%%%%%%%%%%%%%%%%%%%%%%%%%%%%%%

It is useful to obtain the form of the function $f\left(R,\mathcal{R}\right)$ in all the previous cases as the scalar-tensor form of the theory is only a derived form and used to provide an easier platform to perform the calculations. To  obtain $f\left(R,\mathcal{R}\right)$, we are going to use the definition of the potential given by Eq.~\eqref{potdef}.
If we replace here the definitions of the scalar fields given in Eqs.~\eqref{phidef} and~\eqref{psidef}, we obtain a partial differential equation for $f$ as
\be\label{difpot}
V\left(f_R,f_\mathcal{R}\right)=-f\left(R,\mathcal{R}\right)+f_R R+f_\mathcal{R} \mathcal{R},
\ee
where the subscripts $R$ and $\mathcal{R}$ represent derivatives with respect to these variables, respectively.

As we have seen in the previous section the solutions for the scalar field can have zeros and assume a negative sign. While this does not compromise the physical meaning of the solution in the generalized hybrid gravity picture (in this picture the scalar fields are not real matter fields) it can lead to further constraints on the form of the action. One might require that  the derivatives of $f$, which correspond to the scalar fields by Eqs.~(\ref{phidef}) and~(\ref{psidef}), are never zero. In this case the solution we have found will only be meaningful in the intervals in which the scalar field have a fixed sign.

 Let us now analyze this equation for the various forms of the potential obtained so far.

%%%%%%%%%%%%%%%%%%%%%%%%%%%%%%%%%%%%%%%%%%%%%%%%%%%%%%%%%%%%%%%%%%%%
\subsection{The previous solutions pulled into the generalized hybrid gravity}
%%%%%%%%%%%%%%%%%%%%%%%%%%%%%%%%%%%%%%%%%%%%%%%%%%%%%%%%%%%%%%%%%%%%

%%%%%%%%%%%%%%%%%%%%%%%%%%%%%%%%%%%%%%%%%%%%%%%%%%%%%%%%%%%%%%%%%%%%
\subsubsection{The de Sitter solution}\label{method1h}
%%%%%%%%%%%%%%%%%%%%%%%%%%%%%%%%%%%%%%%%%%%%%%%%%%%%%%%%%%%%%%%%%%%%

The potential obtained in Sec.~\ref{method1} is given by $V=12{\Lambda}\left(\varphi-\psi\right)$ and therefore Eq.~\eqref{potdef} can be written as
\be\label{eqdiff1}
f\left(R,\mathcal{R}\right)=f_R \left(R-12{\Lambda}\right)+f_\mathcal{R}\left(\mathcal{R}-12{\Lambda}\right).
\ee
Defining two new variables as
$\bar{R}=R-12{\Lambda}$ and $\bar{\mathcal{R}}=\mathcal{R}-12{\Lambda}$,
the derivatives of the function $f$ remain the same for the new variables and the equation can be written in a simpler manner as
$
f\left(\bar{R},\bar{\mathcal{R}}\right)=f_{\bar{R}}\bar{R}+f_{\bar{\mathcal{R}}}\mathcal{R}
$,
which has a general solution of the form
$
f\left(\bar{R},\bar{\mathcal{R}}\right)=g\left({\bar{\mathcal{R}}}/{\bar{R}}\right)\bar{R}
$,
where $g$ is an arbitrary function. Coming back to the variables $R$ and $\mathcal{R}$,
we obtain
\be\label{sollin}
f\left(R,\mathcal{R}\right)=g\left(\frac{\mathcal{R}-12{\Lambda}}{R-12{\Lambda}}\right)\left(R-12{\Lambda}\right)\,,
\ee
where $g$ is still an arbitrary function.
As an example, let us consider a function $g$ of the form
\be
g\left(\frac{\mathcal{R}-12{\Lambda}}{R-12{\Lambda}}\right)=\exp\left(\frac{R-12{\Lambda}}{\mathcal{R}-12{\Lambda}}\right).
\ee
Inserting this function $g$ into Eq.~\eqref{sollin} to obtain $f$, inserting the result into Eq.~\eqref{difpot} and using the definitions of the scalar fields given by Eqs.~\eqref{phidef} and~\eqref{psidef}, we recover the result $V=12{\Lambda}\left(\varphi-\psi\right)$ for the potential and, consequently, the solutions for the scalar fields are the ones obtained in Sec.~\ref{method1}. 

However, not all forms of the function $g$ can be used to obtain the correct solutions. Take for example a function $g$ of a power-law form, i.e., 
$
g\left(\frac{\mathcal{R}-12{\Lambda}}{R-12{\Lambda}}\right)=\left(\frac{\mathcal{R}-12{\Lambda}}{R-12{\Lambda}}\right)^\gamma
$, where the exponent $\gamma$ is a number.
This form of the function $g$ is compatible with Eq.~\eqref{eqdiff1} but it does not possess the solution we have reconstructed. This is due to the fact that the relationship between the scalar fields and the Ricci scalars given by $\varphi\left(R,\mathcal R\right)$ and $\psi\left(R,\mathcal R\right)$ cannot be inverted to obtain $R\left(\varphi,\psi\right)$ and $\mathcal R\left(\varphi,\psi\right)$. The equivalence between the metric and the scalar-tensor representations breaks down and one cannot immediately say that the theory will have the solutions reconstructed.

%%%%%%%%%%%%%%%%%%%%%%%%%%%%%%%%%%%%%%%%%%%%%%%%%%%%%%%%%%%%%%%%%%%%
\subsubsection{Solution with a simplified potential equation I}
\label{method2h}
%%%%%%%%%%%%%%%%%%%%%%%%%%%%%%%%%%%%%%%%%%%%%%%%%%%%%%%%%%%%%%%%%%%%

The potential obtained in Sec.~\ref{method2} is given by $V=V_0$, a
constant. We can then write Eq.~\eqref{potdef} as
\be
f\left(R,\mathcal{R}\right)=f_R R+f_\mathcal{R} \mathcal{R}-V_0.
\ee
This equation has a general solution of the form
\be\label{solconst}
f\left(R,\mathcal{R}\right)=-V_0+g\left(\frac{\mathcal{R}}{R}\right)R,
\ee
where $g$ is again an arbitrary function.
As an example, let us consider a function $g$ of the form
\be
g\left(\frac{\mathcal R}{R}\right)=\exp\left(\frac{R}{\mathcal R}\right).
\ee
This exponential form allows us to invert the relation between the
scalar fields and the Ricci tensors given by $\varphi\left(R,\mathcal
R\right)$ and $\psi\left(R,\mathcal R\right)$ to obtain
$R\left(\varphi,\psi\right)$ and $\mathcal
R\left(\varphi,\psi\right)$, and therefore we recover the constant
potential $V_0$ and the solutions for the scalar fields presented in
Sec.~\ref{method2}.
Interestingly, in this case
the 
power-law form for the function $g$, i.e., 
$g\left(\frac{\mathcal R}{R}\right)=\left(\frac{\mathcal R}{R}\right)^\gamma$,
also allows us to invert the relation between the scalar fields and the Ricci tensors.

%%%%%%%%%%%%%%%%%%%%%%%%%%%%%%%%%%%%%%%%%%%%%%%%%%%%%%%%%%%%%%%%%%%%
\subsubsection{Solution with a simplified potential equation II}
\label{method3h}
%%%%%%%%%%%%%%%%%%%%%%%%%%%%%%%%%%%%%%%%%%%%%%%%%%%%%%%%%%%%%%%%%%%%

The simple potential obtained in Sec.~\ref{method3} is given by $V=-3\Omega^2\left(\varphi-\psi\right)$, see
Eq.~(\ref{potiispec}). This potential is of the same form of the one obtained in Sec.~\ref{method1} where the only difference is the factor $-3\Omega^2$ instead of $12\Lambda$. Therefore, the procedure to obtain the function $f$ is the same as in Sec.~\ref{method1h} and the solution is
\be
f\left(R,\mathcal{R}\right)=g\left(\frac{\mathcal{R}+6c}{R+6c}\right)\left(R+6c\right).
\ee
where $g$ is an arbitrary function.

%%%%%%%%%%%%%%%%%%%%%%%%%%%%%%%%%%%%%%%%%%%%%%%%%%%%%%%%%%%%%%%%%%%%
\subsubsection{Solution with simplified scalar field equations}
\label{method4h}
%%%%%%%%%%%%%%%%%%%%%%%%%%%%%%%%%%%%%%%%%%%%%%%%%%%%%%%%%%%%%%%%%%%%

The potential obtained in Sec.~\ref{method4} is given by $V=V_0\left(\varphi-\psi\right)^2$. With this potential, Eq.~\eqref{potdef} can be written as
\be
f\left(R,\mathcal{R}\right)=f_R R+f_\mathcal{R} \mathcal{R}-V_0\left(f_R+f_\mathcal{R}\right)^2.
\ee
This equation is a particular case of the Clairaut equation, with a general solution of the form
\be
f\left(R,\mathcal{R}\right)=RC_1+\mathcal{R}C_2-V_0\left(C_1+C_2\right)^2,
\ee
where $C_1$ and $C_2$ are arbitrary constants of integration.

%%%%%%%%%%%%%%%%%%%%%%%%%%%%%%%%%%%%%%%%%%%%%%%%%%%%%%%%%%%%%%%%%%%%
\subsubsection{Solution with nonflat geometry}
\label{method5h}
%%%%%%%%%%%%%%%%%%%%%%%%%%%%%%%%%%%%%%%%%%%%%%%%%%%%%%%%%%%%%%%%%%%%

The potential obtained in Sec.~\ref{method5} is given by $V=V_0$,
a constant potential. This is the same potential as in Sec.~\ref{method2} and therefore the procedure to find the function $f$ is the same and the solution is given by Eq.~\eqref{solconst}.

%%%%%%%%%%%%%%%%%%%%%%%%%%%%%%%%%%%%%%%%%%%%%%%%%%%%%%%%%%%%%%%%%%%%
\subsubsection{Solution with perfect fluid matter}
\label{method6h}
%%%%%%%%%%%%%%%%%%%%%%%%%%%%%%%%%%%%%%%%%%%%%%%%%%%%%%%%%%%%%%%%%%%%

In Sec.~\ref{method6} we presented two different solutions with perfect fluid matter. These solutions were generalizations of the solutions obtained in Secs.~\ref{method1} and~\ref{method2}, where the potentials were given by $V=12{\Lambda}\left(\varphi-\psi\right)$ and $V=V_0$, respectively. Therefore, the solutions for the function $f$ remain the same, given by Eqs.~\eqref{sollin} and~\eqref{solconst} respectively.

%%%%%%%%%%%%%%%%%%%%%%%%%%%%%%%%%%%%%%%%%%%%%%%%%%%%%%%%%%%%%%%%%%%%
\section{Conclusions}\label{conclusion}
%%%%%%%%%%%%%%%%%%%%%%%%%%%%%%%%%%%%%%%%%%%%%%%%%%%%%%%%%%%%%%%%%%%%

In this paper, we devised a number of reconstruction strategies to obtain  new exact cosmological solutions in the context of  the generalized hybrid metric-Palatini gravity. This theory can be recast in a scalar-tensor theory with two scalar fields whose solutions can be computed analytically for a convenient choice of the potential or the scale factor. Using the new methods we obtained a number of physically interesting solutions including  power-law and exponential  scale factors.

These solutions reveal some crucial differences not only with respect to GR but also to $f(R)$ gravity. For example it becomes evident that a given behavior of the scale factors in vacuum has always a counterpart in the presence of a perfect fluid, as shown in Sec.~\ref{method6}. 

The impact in terms of the interpretation of the cosmological dark phenomenology via this class of theories is clear: a negligible matter distribution still gives rise to expansion laws which are compatible with the observations without requiring the total baryonic matter to be a relevant percentage of the total energy density. This is particularly important in the case of the de Sitter solution given in Sec.~\ref{method1}, because it implies that in these theories inflation could start even if matter is not negligible, and in the result of  Sec.~\ref{method4}, which shows that in these theories matter is not central to obtain the classical $t^{2/3}$ Friedmann solution. 

Another interesting aspect of the solutions we found concern the behavior of vacuum cosmology and the role of the spatial curvature. In the first case, see Sec.~\ref{method3}, we obtained that the generalized hybrid metric-Palatini theory can have a surprisingly complex behavior in the flat vacuum case. This has important consequences in terms, for example, of the far future of the cosmological models and the meaning of the cosmic no hair theorem in this framework. 

Like in any higher order model, in the generalized hybrid metric-Palatini theory  the spatial curvature has  an important role. Even small deviations from perfect flatness are able to change dramatically the evolution of the cosmology. The result of Sec.~\ref{method5}  gives us a glimpse of these differences and their magnitude, showing that vacuum closed models will collapse, whereas open ones expand forever both with a relatively simple expansion law. 

It is also interesting to note that our results imply that generalized hybrid metric-Palatini gravity can generate a de Sitter evolution without the appearance of an explicit cosmological constant in the action. Instead, the presence of such a constant leads, throughout our method, to solutions which are radiationlike, see Sec.~\ref{method2}. This result gives us information on the properties of effective dark fluid related to the non Hilbert-Einstein terms present in this theory. It also implies,  by the cosmological no-hair theorem, that these solutions should be unstable. The stability of the other solutions we have found cannot be obtained with the same ease. This is a limitation of the reconstruction in general. It allows to obtain some exact solution but it does not offer information on their stability which has to be investigated with different approaches. 

Our results suggests that the hybrid theory space is a priori large and requires further investigation not only in terms of the exploration of the solution space, but also in terms of the stability of these solutions.
A final thought should be spent on the role of our results for the testability of this class of theory. It is clear that the solutions we have found can be used to perform a number of  cosmological tests, e.g., the ones based on distances, for example. However, a more complete use of the accuracy of the current observations, such as cosmic microwave background anisotropies, requires a full analysis of the cosmological perturbations which is well outside of the scope of our contribution.

%%%%%%%%%%%%%%%%%%%%%%%%%%%%%%%%%%%%%%%%%%%%%%%%%%%%%%%%%%%%%%%%%%%%
\begin{acknowledgments}
JLR acknowledges financial support from Funda\c{c}\~{a}o para a
Ci\^{e}ncia e Tecnologia (FCT) - Portugal for an FCT-IDPASC Grant
No.~PD/BD/114072/2015.  SC is supported
by an Investigador FCT Research contract through project IF/00250/2013
and acknowledges financial support provided under the European Union's
H2020 ERC Consolidator Grant ``Matter and strong-field gravity: New
frontiers in Einstein's theory'' Grant Agreement No.~MaGRaTh646597,
and under the H2020-MSCA-RISE-2015 Grant No.~StronGrHEP-690904.
JPSL acknowledges financial support from
the FCT Project No.~UID/FIS/00099/2013,
the 
FCT Grant
No.~SFRH/BSAB/128455/2017, the Grant
No.~88887.068694/2014-00 from 
Aperfei\c coamento do Pessoal de N\'\i vel Superior (CAPES), Brazil,
and thanks Piotr Chru\'sciel and the 
Gravitational Physics Group at the 
Faculty of Physics, University of Vienna,
for hospitality.
FSNL acknowledges financial support of 
FCT through an Investigador FCT Research contract, with reference
IF/00859/2012 and the Grant No.~PEst-OE/FIS/UI2751/2014.  
\end{acknowledgments}
%%%%%%%%%%%%%%%%%%%%%%%%%%%%%%%%%%%%%%%%%%%%%%%%%%%%%%%%%%%%%%%%%%%%

%%%%%%%%%%%%%%%%%%%%%%%%%%%%%%%%%%%%%%%%%%%%%%%%%%%%%%%%%%%%%%%%%%%%

\end{document}